\documentclass{article}
\usepackage{arxiv}
\usepackage[utf8]{inputenc} % allow utf-8 input
\usepackage[T1]{fontenc}    % use 8-bit T1 fonts
\usepackage{hyperref}       % hyperlinks
\usepackage{url}            % simple URL typesetting
\usepackage{booktabs}       % professional-quality tables
\usepackage{amsfonts}       % blackboard math symbols
\usepackage{nicefrac}       % compact symbols for 1/2, etc.
\usepackage{microtype}      % microtypography
\usepackage{subcaption}
\usepackage{amsmath,amssymb,amsfonts}
\usepackage{multirow}
\usepackage{graphicx}
\usepackage{algorithm}
\usepackage{algorithmic}

\title{Revisiting Representation Learning for Singing Voice Separation with Sinkhorn Distances}

\author{
  Stylianos I. Mimilakis $^{*}$ \\
  Semantic Music Techn. Group\\
  Fraunhofer-IDMT\\
  Ilmenau, Germany\\
  \texttt{mis@idmt.fraunhofer.de} \\
   \And
 Konstantinos Drossos \thanks{Equally contributing authors.} \\
  Audio Research Group\\
  Tampere University\\
  Tampere, Finland \\
  \texttt{konstantinos.drossos@tuni.fi} \\
     \And
 Gerald Schuller \\
  Applied Media Systems Group\\
  Technical University of Ilmenau\\
  Ilmenau, Germany \\
  \texttt{gerald.schuller@tu-ilmenau.de} \\
}

\begin{document}
\maketitle

\begin{abstract}
In this work we present a method for unsupervised learning of audio representations, focused on the task of singing voice separation. We
build upon a previously proposed method for learning representations of time-domain music
signals with a re-parameterized denoising autoencoder,
extending it by using the
family of Sinkhorn distances with entropic regularization. We evaluate our method on the freely available MUSDB18 dataset of professionally produced music recordings, and our results show that Sinkhorn distances with small strength of entropic regularization are marginally improving the performance of informed singing voice separation. By increasing the strength of the entropic regularization, the learned representations of the mixture signal consists of almost perfectly additive and distinctly structured sources.
\end{abstract}

% keywords can be removed
\setcounter{footnote}{0} 
\section{Introduction}
\label{sec:intro}
Recent advances in music source separation rely on deep learning (DL) approaches that can be discriminated in two categories. In the first category the separation approaches operate in the STFT domain~\cite{stoter19, spleeter2019}, and are denoted as spectral-based approaches. In the second category the separation approaches operate directly on the waveform signals~\cite{demucs, meta_tasnet}, i.e., the approaches are trained end-to-end, and are denoted as waveform-based approaches. Spectral and waveform based approaches
have in common that they implicitly compute source-dependent masks that are applied to the mixture signal, prior to the reconstruction of the target signals~\cite{stoter19, spleeter2019, demucs, meta_tasnet}\footnote{Subject to the masking strategy, we refer to the adaptation of Conv-TasNet~\cite{tasnet-19} for music signals also presented in~\cite{demucs}.}. 

Although the implicit masking is shown to be a simple and robust method to learn source dependent patterns for source separation~\cite{mappings2020}, one could expect that waveform based approaches would significantly outperform the spectral ones. That is because waveform based approaches are optimized using time-domain signals that also contain the phase information, that unarguably carries important signal information~\cite{cano13:rethinking, cano2019:MSS, magron:2018:interspeech} and has been neglected by many spectral based approaches~\cite{stoter19, spleeter2019, drossos18, mim18}. Nonetheless, previously conducted experiments and reported results suggest that spectral based approaches have comparable or marginally better separation performance to the waveform ones~\cite{demucs, meta_tasnet, stoter19}. Since both waveform and spectral approaches rely on DL and for both approaches a considerable engineering effort has been directed to the employed neural architecture, it is evident that the difference in the performance between the two different approaches can be attributed to the utilized signal representation that is used for separation. 

For the spectral-based approaches the utilized representation is the non-negative signal representation offered by the magnitude of the STFT. For the waveform-based approaches the representation is computed by trainable encoding functions, commonly neural networks. The parameters of the encoding functions are optimized jointly with the rest of the separation model. The optimization of the separation model, and thus also the encoding functions that compute the representations, is based on minimizing loss objectives that assess the reconstruction of the signals of the target sources given the mixture signal as input~\cite{demucs, meta_tasnet}. In this case and subject to the representations, the learning is performed using solely discriminative optimization objectives, that aim at distinguishing between the mixture and the target sources. As shown in~\cite{score_matching_daes}, this could potentially impose severe limitations in the generalization capabilities of the learned representations, as the learning process based on discriminative objectives does not aim at capturing the essential structure of the signals~\cite{DAEs:Vincent-2008, DAEs:Vincent-2010}. Furthermore, the learned representations obtained by approaches utilizing end-to-end training are not easily nor intuitively interpreted, compared to the pre-computed signal representations that utilize the STFT.

In an attempt to learn music signal representations that capture the structure of the music signals, are interpretable, and consist of attributes that are useful for music source separation, the focus is given on neural-based representation learning~\cite{simply_bengio_rl}. The following sections present a new and simple method for learning representations of time-domain music signals. The proposed method is characterized as unsupervised because the optimization of the method does not depend on labelled categorical data, i.e., labels for distinguishing between the music sources, and the representation attributes (discussed later) are learned using unsupervised training objectives.
Furthermore, these training objectives do not target the learning of the unmixing function, i.e., the mapping from the mixture to the target source signal. This in turn, alleviates the need of having either labelled or paired training data (i.e., matched multi-track audio data of each corresponding source). However, the proposed method still requires isolated source's audio signals, but this information is more accessible than paired multi-track data. 

The rest of this manuscript is organized as follows: Section~\ref{ch:rl:sec:sota} provides information regarding previously published research that is related to representation and interpretable representation learning for audio and speech processing and enhancement. The proposed method for learning representations is described in Section~\ref{ch:rl-sec-pm}, followed by the experimental procedure described in Section~\ref{ch5:sec:exp_proc}. Section~\ref{ch5:sec:results} presents and discusses the results obtained from the experimental procedure, including visualizations of the obtained representation(s). Section~\ref{ch5:sec:summary} summarizes the findings presented in this manuscript.

\subsection*{Notation}
Bold lowercase letters, e.g., ``$\mathbf{x}$'', denote vectors and bold uppercase letters, e.g. ``$\mathbf{X}$'', denote matrices. The $l$-th element of a vector is denoted as $\mathrm{x}_{[l]}$. Similarly, matrix elements are denoted as $\mathrm{X}_{[l, l']}$. The element-wise multiplication and division are denoted as ``$\odot$'' and ``$\oslash$'', respectively.

\section{Related Work}\label{ch:rl:sec:sota}
The proposed method is based on the denoising autoencoder (DAE) model~\cite{DAEs:Vincent-2010} that can be used also for unsupervised learning of signal features and   representations, other than trained in a supervised way to separate music sources~\cite{mappings2020}. The DAE can efficiently learn the empirical distribution of the signal of interest, i.e., the signal to be denoised~\cite{DAEs:Vincent-2010, score_matching_daes}. This is achieved by optimizing the DAE with the unsupervised objective to reconstruct the clean signal from a noisy version. Subject to this work, the underlying assumption is that a model that efficiently learns the empirical distribution, it can be employed to compute signal representations that can be used to characterize the clean signal of interest. Contrary to the DAE, the proposed method uses a simple and real-valued sinusoidal-based model for the \emph{decoding} functions. The sinusoidal model consists of amplitude-modulated cosine functions, and whose parameters are jointly optimized with the rest of the DAE. The motivation behind using a sinusoidal model as a decoding function is to guide via back-propagation the encoding layers of the DAE to learn and convey information regarding the energy of specific cosine functions that compose the audio signal. This leads to interpretable representations akin to the STFT.

Employing a vastly used digital signal processing operation for decoding functions is inspired by two works. The first work introduces the concept of differentiable digital signal processing~\cite{ddsp} where the parameters of common digital signal processing functions are optimized by means of back-propagation. In the case of the proposed method, back-propagation is applied with respect to the parameters of a simple signal model that is based on sinusoidal functions. The second work that the proposed method is inspired from, is the Sinc-Network presented in~\cite{sinc_net}. The Sinc-Network uses sinc functions in the \emph{encoding} layers of convolutional kernels for interpretable deep learning. The Sinc-Network has been extended to complex-valued representations for speaker separation~\cite{filterbank_design_e2e}. 

The proposed method differs from~\cite{filterbank_design_e2e} as the representation of the proposed method is real-valued, alleviating the cumbersome signal processing operations on complex numbers.
Furthermore, the proposed method differs from approaches that initialize the front-end parts of the networks with cosine functions~\cite{adaptive_fe_ss} that are then updated by means of back-propagation. The difference is that the proposed method inherits the cosine functions as a part of the model to be optimized. Finally,
the proposed method is similar to the sound source separation method presented in~\cite{tzinistwostep}. In~\cite{tzinistwostep} an encoder gets as an input the signals of the sources and their corresponding mixture, and outputs latent representations of the signals of each source and the mixture. Then, using these latent representations, the method calculates and applies source dependent masks to the latent representation of mixture. The result of the application of the masks is given as an input to the decoder, which outputs an estimation of the signal of each source. The encoder and the decoder are jointly optimized to minimize the reconstruction error between the ground truth and the estimated signals of each source, i.e., a discriminative training is performed. However, using reconstruction objectives in a discriminative setting for separating only specific sources, could severely restrict the representation learning capabilities of encoder-decoder methods~\cite{score_matching_daes}. In contrast, the proposed method uses information from the mixture and target source signals using unsupervised and non-discriminative training objectives that aim at capturing the structure of the music signals.

\section{Proposed Method}\label{ch:rl-sec-pm}
The proposed method employs an encoder $E(\cdot)$ and a decoder $D(\cdot)$. The input to the method is a music signal, $\mathbf{x} \in \mathbb{R}_{[-1,1]}^{N}$, with $N$ time-domain samples. The output of the method is the learned non-negative representation of $\mathbf{x}$, $\mathbf{A} \in \mathbb{R}_{\geq 0}^{C \times T}$, with $T$ templates of $C$ features. The $C$ features can be viewed as analogous to the frequency bins and the $T$ templates as the analogous to the time-frames in a time-frequency representation. $\mathbf{A}$ is computed by the encoder $E(\cdot)$, and is interpreted as the magnitude information for a real-valued, sinusoidal-based model,
employed by the decoder $D(\cdot)$.

To optimize $E(\cdot)$, the decoder $D(\cdot)$ is used. In addition to this, a data-set of monaural (single channel) recordings of singing voice, $\mathbf{x}_{\text{v}}\in\mathbb{R}_{[-1,1]}^{N}$, and accompanying musical instruments $\mathbf{x}_{\text{ac}}\in\mathbb{R}_{[-1,1]}^{N}$ is used. Using $\mathbf{x}_{\text{v}}$ two synthetic signals are created. The first synthetic signal, $\tilde{\mathbf{x}}_{\text{m}} \in \mathbb{R}_{[-1,1]}^{N}$, is the result of an additive corruption process, where the accompanying musical instruments such as drums, guitars, synthesizers, and bass (i.e. a generic multi-modal distribution-based noise) are added to $\mathbf{x}_{\text{v}}$: \[ \tilde{\mathbf{x}}_{\text{m}} = \mathbf{x}_{\text{v}} + \mathbf{x}_{\text{ac}} \text{ .}\]
The second synthetic signal, $\tilde{\mathbf{x}}_{\text{v}}\in \mathbb{R}_{[-1,1]}^{N}$, is also the result of a corruption process, where Gaussian noise is added to $\mathbf{x}_{\text{v}}$, independently of the amplitude of ${\mathbf{x}}_{\text{v}}$.

\begin{figure}[!ht]
    \centering
    \includegraphics[width=0.96\columnwidth, keepaspectratio]{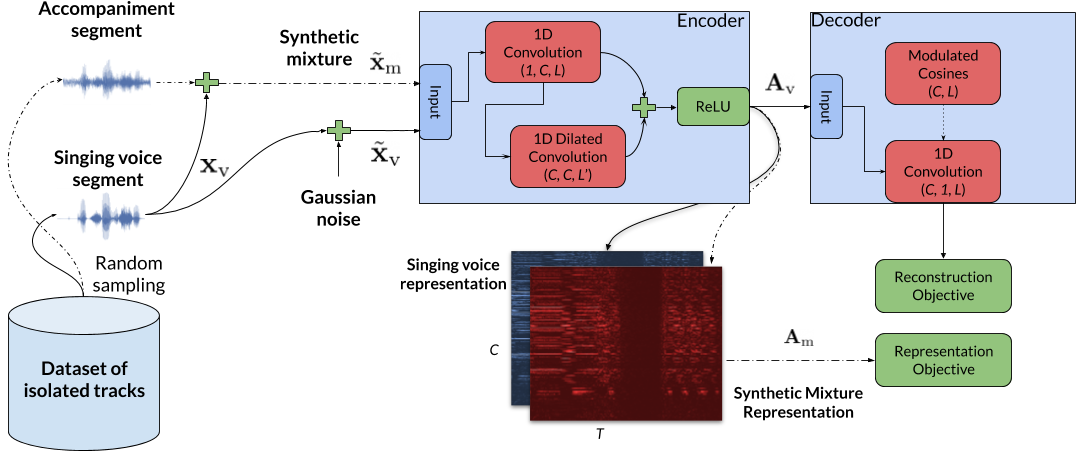}
    \caption{Overview of the proposed method for representation learning.}
    \label{fig:method-icml}
\end{figure}

During training, the encoder $E(\cdot)$ computes two non-negative representations $\mathbf{A}_{\text{m}} \in \mathbb{R}_{\geq 0}^{C \times T}$ and $\mathbf{A}_{\text{v}} \in \mathbb{R}_{\geq 0}^{C \times T}$, using the two above mentioned synthetic signals, $\tilde{\mathbf{x}}_{\text{m}}$ and $\tilde{\mathbf{x}}_{\text{v}}$, respectively. $\mathbf{A}_{\text{v}}$ is used as input to $D(\cdot)$, and $D(\cdot)$ outputs an approximation of the clean singing voice signal $\mathbf{x}_{\text{v}}$, denoted as $\hat{\mathbf{x}}_{\text{v}}$. $\mathbf{A}_{\text{m}}$ is solely used to calculate an additional loss function. This is done in order to allow $E(\cdot)$ to learn information regarding the additive multi-modal noise. An illustration of the training procedure is shown in Figure~\ref{fig:method-icml}.

After the training process of the proposed method, $E(\cdot)$ can take as an input any musical signal $\mathbf{x}$, and will output the representation of $\mathbf{x}$, denoted as $\mathbf{A}$. Furthermore, an approximation of the signal $\mathbf{x}$ can be computed using the decoder $D(\cdot)$. The benefits for doing so, is that $\mathbf{A}$ has good music signal representation attributes that include interpretability, non-negativity, and structured spectrogram-like representations. Furthermore, the inputted signal $\mathbf{x}$ can be approximated from $\mathbf{A}$ using $D(\cdot)$, with a small reconstruction error for the parts that the singing voice signal is active. Consequently, the method could be effectively used in the downstream task of singing voice separation, but it is not limited to.

\subsection{The Encoder}
The encoder $E(\cdot)$ computes the representation(s) using two one-dimensional (1D) convolutions with strides and in series. The first 1D convolution uses a stride size $S$ and a set of $C$ number of kernels, $\mathbf{k}_{c}\in\mathbb{R}^{L}$, where $c = [0,1,\ldots C-1]$ is the kernel index. The temporal length of each kernel $\mathbf{k}$ is $L$ samples. As input to the first convolution, the signals $\tilde{\mathbf{x}}_{\text{m}}$ and $\tilde{\mathbf{x}}_{\text{v}}$ are used. The outputs are the \emph{latent} representations $\tilde{\mathbf{H}}_{\text{m}}\in\mathbb{R}_{\geq0}^{C\times T}$ and $\tilde{\mathbf{H}}_{\text{v}}\in\mathbb{R}_{\geq0}^{C\times T}$, respectively. More formally, the latent representations are computed as
\begin{align}\label{eq:e-conv1_xm}
    \mbox{$\tilde{\mathrm{H}}_{\text{m}}$}_{[c, t]} = \sum\limits_{l=0}^{L-1} \mbox{${{\tilde{\mathrm{x}}}_{\text{m}}}$}_{[St + l]}\,{\mathrm{k}_{c}}_{[l]}\\
    \label{eq:e-conv1_xv}
    \mbox{$\tilde{\mathrm{H}}_{\text{v}}$}_{[c, t]} = \sum\limits_{l=0}^{L-1} \mbox{${{\tilde{\mathrm{x}}}_{\text{v}}}$}_{[St + l]}\,{\mathrm{k}_{c}}_{[l]}\text{,}
\end{align}
where $t \in [0,1,\ldots, T-1]$ and $l \in [0,1,\ldots L-1]$ are integers denoting the time-frame and the kernel sample indices, respectively. Appropriate zero-padding is applied to $\tilde{\mathbf{x}}_{\text{m}}$ and $\tilde{\mathbf{x}}_{\text{v}}$, so that $T=\lceil N/S \rceil$. %and $\lceil\cdot\rceil$ is the ceiling function. 
Each latent representation is used as an input to the second 1D convolution, which uses another set of
$C$ kernels, $\mathbf{K}'_{c'}\in\mathbb{R}^{L' \times C}$, with a temporal length of $L'$ samples, that is $L' << L$. The output channels are indexed by $c'$, where $c'=[0,1,\ldots,C-1]$.

The outputs of the second convolution using the previously computed representations, are denoted by  ${\mathbf{H}}_{\text{m}}\in\mathbb{R}^{C \times T}$ and 
${\mathbf{H}}_{\text{v}}\in\mathbb{R}^{C \times T}$, respectively. The second 1D convolution is performed with a dilation factor of $\phi$ and a unit stride $S=1$, as
\begin{align}\label{eq:e-conv2}
    {\mathrm{H}_{\text{m}}}_{{[c', t]}} = \sum\limits_{c=0}^{C-1} \sum\limits_{l'=0}^{L'-1}
    \mbox{$\tilde{\mathrm{H}}_{\text{m}}$}_{[c, t + \phi l']} {\mathrm{K}^{'}_{c'}}_{[l', c]} \\
    {\mathrm{H}_{\text{v}}}_{{[c', t]}} = \sum\limits_{c=0}^{C-1} \sum\limits_{l'=0}^{L'-1}
    \mbox{$\tilde{\mathrm{H}}_{\text{v}}$}_{[c, t + \phi l']} {\mathrm{K}^{'}_{c'}}_{[l', c]}
    \, .
\end{align}

Then, the representations $\mathbf{A}_{\text{m}}$ and $\mathbf{A}_{\text{v}}$ are computed using the previously computed latent representations, $\mathbf{H}_{\text{m}}$ and $\mathbf{H}_{\text{v}}$ respectively, and by means of residual connections, followed by the application of the rectified linear unit (ReLU) function~\cite{relu_rbm} as
\begin{align}
    \mathbf{A}_{\text{m}} &= \text{ReLU}(\mathbf{H}_{\text{m}} + \tilde{\mathbf{H}}_{\text{m}}) \\
    \mathbf{A}_{\text{v}} &= \text{ReLU}(\mathbf{H}_{\text{v}} + \tilde{\mathbf{H}}_{\text{v}}) \text{ . }
\end{align}
\noindent
The application of the ReLU function promotes non-negative and sparse representations by preserving positive values and setting the rest to zero~\cite{papyan17}, and is shown to be particularly useful in general modelling of audio signals~\cite{non_negative_paris}.
Another targeted and useful attribute of the learned representation is that of smoothness~\cite{adaptive_fe_ss, non_negative_paris}, especially useful when real-valued cosine functions are involved in auto-encoding or separation models~\cite{adaptive_fe_ss}. Smoothness refers to the slow time variation of the representation, and is useful for general audio signal modelling. That is because the modelling of audio signals based on cosine functions requires the phase information for reconstruction. Phase information is usually encoded as the sign (positive or negative value) of the real-valued representation, that varies along the time-frames of the representation. Since the negative values are nullified by the application of the ReLU function, neighbouring time-frames, that convey similar information for music signals are expected to be non-smooth. To compensate for the expected non-smoothness, the second convolution operation uses dilated convolutions that aggregate temporal information from neighboring time-frames~\cite{dilated_convs} and updated using residual connections.

In order to enforce the learning of smooth representations, a representation objective is introduced. The introduced objective is a loss function that the encoder has to minimize. The most straightforward way to enforce the smoothness is to compute the norm of the first-order differences of the representation~\cite{tv_loss}. To do so, the (anisotropic) \index{total-variation denoising}{total-variation denoising} loss is used. Specifically, the representation of $\tilde{\mathbf{x}}_{\text{m}}$, ${\mathbf{A}}_{\text{m}}$, is used to compute the total variation denoising ($\mathcal{L}_{\text{TV}}(\cdot)$) as
\begin{align}
    \label{eq:tv_loss}
    {\mathcal{L}_{\text{TV}}}(\mathbf{A}_\text{m}) &= \frac{1}{CT} \Big( \sum_{c=1}^{C-1}\sum_{t=0}^{T-1}|{\mathrm{A}_\text{m}}_{[c, t]} - {\mathrm{A}_\text{m}}_{[c-1,t]}|\nonumber\\
    &+ \sum_{t=1}^{T-1}\sum_{c=0}^{C-1} |{\mathrm{A}_\text{m}}_{[c, t]} - {\mathrm{A}_\text{m}}_{[c, t-1]}|\Big)\text{.}
\end{align}
Practically, $\mathcal{L}_{\text{TV}}(\cdot)$ penalizes $E(\cdot)$ by the norm of the first order difference across both time-frames $T$ and templates $C$. The former promotes slow time varying representations as the magnitude of the STFT representation, and the latter promotes a grouping of the template activity. The previously mentioned attributes of the desired representation attributes are formed from domain knowledge that is based on the STFT. Furthermore, $\mathcal{L}_{\text{TV}}(\cdot)$ is an unsupervised objective that depends only on the representation $\mathbf{A}_{\text{m}}$.

Although $\mathcal{L}_{\text{TV}}(\cdot)$ seems an attractive loss function due to its simple computation, it has a severe limitation. According to~\cite[Theorem~2]{wasserstein_gan} the total-variation distance, and in this particular case the sum of absolute differences employed in Eq.\eqref{eq:tv_loss}, is not a suitable loss function for data distributions supported by low-dimensional manifolds. Instead, optimal transportation distances are more suitable. Under the hypothesis that both the singing voice and the mixture signals, and their corresponding representations can be described by low-dimensional manifold(s), an alternative unsupervised objective to $\mathcal{L}_{\text{TV}}(\cdot)$ is also examined. 

\index{Sinkhorn distances}{Sinkhorn distances} $\mathcal{L}_{\text{SK}}(\cdot)$ allow an efficient computation of optimal transportation loss~\cite{sinkhorn}. More specifically and subject to the goal of this work, Sinkhorn distances are computed as
\begin{equation}\label{eq:sk-dist}
    \mathcal{L}_{\text{SK}}(\mathbf{A}_{\text{m}}) = \langle\mathbf{P}_{\lambda}, \psi(\mathbf{A}_{\text{m}}) \rangle \text{ , }
\end{equation}
where ``$\langle \cdot, \cdot \rangle$'' is the Frobenious dot-product and $\psi:\mathbb{R}_{\geq 0}^{C \times T} \mapsto \mathbb{R}_{\geq 0}^{T \times T}$ is a function that computes the matrix $\mathbf{M} \in \mathbb{R}_{\geq 0}^{T \times T}$ of pair-wise distances, i.e., 
$ \mathbf{M} = \psi(\mathbf{A}_{\text{m}})$. More specifically, the pair-wise distances are computed as
\begin{equation}\label{eq:pairwise-distance}
 \mathrm{M}_{[t\, , t']} = 
 \Big(\sum_{c=0}^{C-1}(|{\mathrm{A}_{\text{m}}}_{[c, t]} - {\mathrm{A}_{\text{m}}}_{[c, t']}|)^{p}\Big)^{1/p}
\text{. }
\end{equation}
In Eq.\eqref{eq:pairwise-distance} $t,t' \in [0, \ldots, T-1]$ are indices that are used to compute the pair-wise distance between the time-frames T of the representation. Furthermore, $p=1$ is considered in computing the distances. It should be denoted, that 
only for, and prior to, the computation of the loss matrix $\mathbf{M}$, the representation ${\mathbf{A}_{\text{m}}}$ is normalized so that the sum of the features at each time-frame $t$ sum up to unity. More formally, the normalized representation $\mathbf{A}^{\mathit{o}}_{\text{m}}$ is computed as
\[ {\mathrm{A}^{\mathit{o}}_{\text{m}}}_{[c,t]} = 
\frac{{\mathrm{A}_{\text{m}}}_{[c,t]}}{\sum\limits_{c}({\mathrm{A}_{\text{m}}}_{[c, t]} + \frac{1}{C})}
\text{ . }\]
This is done in order to treat ${\mathbf{A}_{\text{m}}}$ as a probability simplex in which the computation of the optimal transportation loss can be computed.

In Eq.\eqref{eq:sk-dist}, $\mathbf{P}_{\lambda} \in \mathbb{R}_{\geq 0}^{T \times T}$ is the transportation plan that is computed by solving the following minimization problem
\begin{equation}\label{eq:argmin_P}
\mathbf{P}_{\lambda} = \underset{\mathbf{P} \in \mathbb{U}(r, c)}{\operatorname*{arg\,min}} \langle \mathbf{P}, \psi(\mathbf{A}_{\text{m}}) \rangle - \frac{1}{\lambda} H(\mathbf{P}) \text{ . }
\end{equation}
In the above minimization problem, $\lambda > 0$ is a scalar the controls the strength of the entropic regularization, and $H(\cdot)$ denotes the entropy function that is computed as\index{entropy function} \[H(\mathbf{P}) = - \sum_{t,t'=0}^{T-1} \mathrm{P}_{[t,t']}\log(\mathrm{P}_{[t,t']}) \text{ . }\] In addition to this, $\mathbb{U}(r,c)$ is the set of non-negative matrices of size $T \times T$ whose rows and columns sum up to $r$ and $c$, respectively. It is further assumed that $r=c=1$. 
For solving the minimization problem of Eq.\eqref{eq:argmin_P} the proposed method for representation learning employs the algorithm presented in~\cite{sinkhorn} that is based on the Sinkhorn-Knopp iterative matrix scaling operator~\cite{sink_iterative} and is pseudo-algorithm is given in Algorithm~\ref{alg:sinkhorn}. In Algorithm~\ref{alg:sinkhorn},  $||\cdot||_{p}$ is the $p$-th vector or matrix norm.
\begin{algorithm}[!t]\index{Sinkhorn-Knopp algorithm}
\caption{Computation of the transportation plan, using Sinkhorn-Knopp's iterative matrix scaling operation~\cite{sink_iterative, sinkhorn}}
\begin{algorithmic}[1]
    \REQUIRE{loss matrix $\mathbf{M}$, entropic regularization scalar $\lambda$, dimensionality $T$, vector of ones $\mathbf{1}_{T}$, number of iterations $iter$, termination threshold $\tau$}
    \STATE Initialize: $K = \text{exp}(-\lambda \mathbf{M})$, $\mathbf{u} = \mathbf{1}_{T} \oslash T$, $\mathbf{v} = \mathbf{1}_{T} \oslash T$, $\mathbf{K}_{\mathbf{u}} = \text{diag}(\mathbf{u})\, \mathbf{K}$\\
    \FORALL{$iter$}
    \STATE $\mathbf{v} \gets T \oslash (\mathbf{K}^{T} \mathbf{u})$
    \STATE $\mathbf{u} \gets 1 \oslash (\mathbf{K}_{\mathbf{u}} \, \mathbf{v})$
    \STATE $o = ||\text{diag}(\mathbf{u}) \, \mathbf{K} \, \text{diag}(\mathbf{v})||_1$
    \IF{$|| o - (\mathbf{1}_{T}\oslash T)||_{1}^{2} < \tau$}
    \STATE stop iterating
    \ENDIF
    \ENDFOR
    \STATE $\mathbf{P}_{\lambda} = \text{diag}(\mathbf{u}) \, \mathbf{K} \, \text{diag}(\mathbf{v})$
    \RETURN Transportation plan $\mathbf{P}_{\lambda}$\\
\end{algorithmic}
\label{alg:sinkhorn}
\end{algorithm}

In Eq.\eqref{eq:tv_loss} and Eq.\eqref{eq:argmin_P} only the representation $\mathbf{A}_\text{m}$ is used to compute the corresponding loss functions. This is performed in order to enforce the encoder $E(\cdot)$ to yield smooth representations on the most realistic corruption scenario. This scenario is the additive generic multi-modal distribution-based noise $\tilde{\mathbf{x}}_{\text{m}}$ that contains also the information regarding the singing voice signal $\mathbf{x}_{\text{v}}$. Thus, the smoothness for the representation of the singing voice is implicitly enforced.

\subsection{The Decoder}
The decoder $D(\cdot)$ takes as an input the representation $\mathbf{A}_{\text{v}}$ and yields an approximation of the clean singing voice signal $\mathbf{x}_{\text{v}}$, denoted as $\hat{\mathbf{x}}_{\text{v}} \in \mathbb{R}_{[-1,1]}^{N}$. Specifically, $D(\cdot)$ models the clean singing voice as a sum of $C$ modulated sinusoidal components that overlap in $\mathbb{R}^{N}$. The components are computed using 1D transposed convolutions with $S$ strides and another set of $C$ number of kernels, $\mathbf{w}_{c}\in\mathbb{R}^{L}$, as
%\begin{align}
%\mbox{$\hat{\mathrm{x}}_{\text{v}}$}_{[St + l]} =& \mathrm{\eta} + \sum_{c=0}^{C-1}\mathrm{A}_{\text{v}_{[c, t]}} {\mathrm{w}_{c}}_{[l]} \text{, where}\label{eq:conv_synthesis}\\
%\mathrm{\eta} =&
%    \begin{cases}
%      0, & \text{if}\ t=0 \\
%      \mbox{$\hat{\mathrm{x}}_{\text{v}}$}_{[S(t-1) + l]}, & \text{otherwise}
%    \end{cases}\label{eq:eta} .
%\end{align}
\begin{equation}
\mbox{$\hat{\mathrm{X}}_{\text{v}}$}_{[l, t]} = \sum_{c=0}^{C-1}\mathrm{A}_{\text{v}_{[c, t]}} {\mathrm{w}_{c}}_{[l]} \text{ , }
\label{eq:conv_synthesis} 
\end{equation}
where ${\hat{\mathbf{X}}}_{\text{v}} \in \mathbb{R}^{L \times T}$ is the matrix containing the modulated components that are used to compute $\hat{\mathbf{x}}_{\text{v}}$ as
\begin{equation}
\mbox{$\hat{\mathrm{x}}_{\text{v}}$}_{[n]} = \sum_{t} \mbox{$\hat{\mathrm{X}}_{\text{v}}$}_{[n - t \, S, t]} \, \forall{n} \in [0, 1, \ldots, N-1] \text{.}\label{eq:eta}
\end{equation}
Eq.~\eqref{eq:eta} is the overlap-add process and follows the assumption that \[
\mbox{$\hat{\mathrm{X}}_{\text{v}}$}_{[n - t \, S, t]} = 0 \, \, \text{ if } (n - t \, S) \not\in [0,1, \ldots, L-1].
\]
.

Similar to the Sinc-Net~\cite{sinc_net} and it's complex-valued extension for speech enhancement~\cite{filterbank_design_e2e}, the proposed method does not allow each $\mathbf{w}_{\text{c}}$ to be updated directly using back-propagation. Instead, each $\mathbf{w}_{\text{c}}$ is re-parameterized by amplitude modulated sinusoidal functions. The back-propagation is computed with respect to the corresponding parameters of the modulated sinusoidal functions. More specifically, each $\mathbf{w}_{\text{c}}$ is computed using \index{re-parameterization with modulated cosines}
\begin{equation}
    \label{eq:cos_reparameterization}
    {\mathrm{w}_{c}}_{[l]} = \text{cos}(2 \pi f_{c}^{2}  \, l + \rho_{c}) \,  {\mathrm{b}_{c}}_{[l]}\text{ ,}
\end{equation}
\noindent
where $\text{cos}(\cdot)$ is the cosine function and $l=[0, \ldots, L-1 ]$ is the time index. The parameters that are jointly learnt with the parameters of the encoder $E(\cdot)$, are the sampling-rate-normalized carrier frequency $f_{c}$, the phase $\rho_{c}$ (in radians), and the modulating signal $\mathbf{b}_{c}\in \mathbb{R}^{L}$. The direct access to natural quantities like the above described, significantly boosts the interpretability of the representation learning method. Additionally, $\mathbf{w}_{c}$ can be sorted according to the carrier frequency $f_c$, promoting intuitive representations. The non-linear squaring operation applied to $f_{c}$ is motivated by experimental results presented in Section~\ref{ch5:sec:results-design-choices}.

There are three reasons for using modulated cosine functions for decoding $\mathbf{A}_{\text{v}}$: a) cosine functions promote interpretability~\cite{sinc_net}, i.e. the representation $\mathbf{A}_{\text{v}}$ is expected to convey amplitude related information for driving a well established synthesis model based on sinusoidal functions~\cite{serra_sms}, b) the auto-encoding operation shares many similarities with the STFT yet without having to deal directly with the phase information, for which supervised based separation works remarkably well~\cite{stoter19, spleeter2019}, and c) amplitude modulations allow an extra degree of freedom in reconstructing signals that cannot be described by pure sinusoidal functions~\cite{serra_sms}. The latter statement is supported by the convolution theorem which states that the element-wise product of two vectors can be expressed in the Fourier domain as their corresponding convolution. Since in the proposed re-parameterization scheme (i.e. Eq.~\eqref{eq:cos_reparameterization}) one of the signals is a cosine function, then $\mathbf{b}_{c}$ is expected to convey timbre information regarding the signal $\mathbf{x}_{\text{v}}$ that was used to compute the reconstruction objective.

After the reconstruction of $\hat{\mathbf{x}}_{\text{v}}$, the negative signal-to-noise ratio (neg-SNR) \cite{uni_ass}, is computed as
\begin{equation}
    \label{eq:neg-snr}
    \mathcal{L}_{\text{neg-SNR}} (\mathbf{x}_{\text{v}}, \hat{\mathbf{x}}_{\text{v}}) = - 10 \, \text{log}_{10}\Big(\frac{||\mathbf{x}_{\text{v}}||_{2}^2}{||\mathbf{x}_{\text{v}} - \hat{\mathbf{x}}_{\text{v}}||_{2}^2}\Big) \text{ , }
\end{equation}
where $||\cdot||_2$ is the $\ell_2$ vector norm, and the negative sign is used to cast the logarithmic SNR as a minimization objective. Then, the overall overall minimization objective for $E(\cdot)$ and $D(\cdot)$ is computed using $\mathcal{L}_{\text{TV}}(\cdot)$ as
\begin{equation}
    \label{eq:tot-loss-TV}
    {L}_{A} = \mathcal{L}_{\text{neg-SNR}}(\mathbf{x}_{\text{v}}, \hat{\mathbf{x}}_{\text{v}}) + \omega \, \mathcal{L}_{\text{TV}}(\mathbf{A}_{\text{m}})\text{,}
\end{equation}
or using $\mathcal{L}_{\text{SK}}(\cdot)$ as
\begin{equation}
    \label{eq:tot-loss-SK}
    {L}_{B} = \mathcal{L}_{\text{neg-SNR}}(\mathbf{x}_{\text{v}}, \hat{\mathbf{x}}_{\text{v}}) +  \omega \, \mathcal{L}_{\text{SK}}(\mathbf{A}_{\text{m}}) \text{, } 
\end{equation}
\noindent
where $\omega$ is a scalar that weights the impact of the representation objective (either $\mathcal{L}_{\text{TV}}(\cdot)$ or $\mathcal{L}_{\text{SK}}(\cdot)$) in the gradient (learning signal) used for optimizing $E(\cdot)$. In addition to this, $L_{A}$ and $L_{B}$ are scalar values that contain the overall loss that is used for optimizing the encoder and the decoder. 
The decoder $D(\cdot)$ computes $\hat{\mathbf{x}}_{\text{v}}$ only from the signing voice representation $\mathbf{A}_{\text{v}}$. That is because it is aimed at learning general representations in an unsupervised and non discriminative fashion. To achieve that by means of the DAE model~\cite{DAEs:Vincent-2010}, it is reasonably assumed that the distribution of the corruption process is constant for all segments in the data-set~\cite{score_matching_daes}. This cannot be assumed for music signal mixtures, as even the distribution of the accompaniment instruments can vary dramatically from one segment to another. Consequently, by making such an assumption it could lead to degenerate representations for singing voice.

\section{Experimental Procedure}\label{ch5:sec:exp_proc}
\subsection{Data-set}
For training and testing the representation learning method the MUSDB18 data-set~\cite{musdb18} is used. The data-set consists of 150 two-channel professionally produced multi-tracks, i.e, the stereophonic signals of bass, drums, singing voice, and other music instruments, that comprise a music mixture. Every signal is sampled at $44100$ Hz.
The multi-tracks are split into training (100 multi-tracks) and testing (50 multi-tracks) subsets.

\subsection{Initialization \& Hyper-parameter Selection}
\subsubsection*{Initialization}
Before the training process, the kernels in the first convolutions are randomly initialized with values drawn from a uniform distribution. 
The bounds of the uniform distribution are $(-\sqrt{\frac{3}{C}}, \sqrt{\frac{3}{C}})$, following the initialization strategy presented in~\cite{kaiming}.
For the decoder, the phase values $\rho_c$ are initialized to zero, and all the elements of the modulating vectors $\mathbf{b}_{c}$ are initialized to the value of $\frac{1}{C+L}$. The initialization of the normalized frequencies $f_c$ is inspired by~\cite{sinc_net} and is performed by first computing the center frequencies of the Mel scale, denoted as  $f_{\text{Mel}}$, in the range of $f_{\text{Hz}}\in [30, \ldots, 22050]$ Hz with a step-size equal to $C$. Then, $f_{\text{Mel}}$ is computed as
\[f_{\text{Mel}} = 2595 \,  \text{log}_{10}(1 + \frac{f_{\text{Hz}}}{700})\]
and the initial value for each component $f_c$ is computed as
\[f_{c} =  \frac{700 \, 10^{\frac{f_{\text{Mel}}}{2595}} - 1}{44100} \text{ . }
\]
\subsubsection*{Hyper-parameter Selection}
For selecting the  hyper-parameters of the convolutional networks and the training procedure, a pilot experiment is conducted. During this experiment, $20$ randomly selected tracks from the training sub-set of MUSDB18 data-set were used. The objective of the pilot experiment is to determine the learning rate and the batch size of the solving algorithm, the standard deviation of the additive Gaussian noise for the corruption processes, described in Section~\ref{ch:rl-sec-pm}, and the convolutional hyper-parameters. To that aim, the proposed method was trained \emph{without} the representation objective, with the only objective to reconstruct the singing voice signal from it's corrupted version. The results from each experimental run were assessed by means of informal listening tests, focusing on the subjective quality of the reconstruction of the singing voice.

The results from the above described experimental procedure are the usage of the adam algorithm~\cite{adam}, with a learning rate equal to $1e^{-4}$ and a batch size of $8$. In addition to this, the following hyper-parameters for the convolutional layers: (number of kernels) $C=800$, (stride size) $S=256$, (temporal length of each kernel in the first encoding layer) $L=2048$, (temporal length of each kernel in the second encoding layer) $L'=5$, and (dilation factor for the second encoding layer) $D=10$ provided perceptually good reconstruction. Furthermore, it was observed that the method converges fast, so for the complete experimental procedure the total number of iterations throughout the whole data is set to $10$. In similar vein, a standard deviation of $1e^{-4}$ for the additive Gaussian noise was found to yield good and relatively fast results from a range of values $[1e^{-5}, 5e^{-5}, 1e^{-4}, 5e^{-4}, 1e^{-3}, 5e^{-3}, 1e^{-2}]$. Based on the available computational resources\footnote{An Nvidia GTX 1050Ti GPU with 6GB of memory.} each multi-track is partitioned in segments of $N=44100$ samples (1 second long).

\subsection{Training}\label{ch5:sec:exp_proc_training}
During training, a set of four multi-tracks is sampled. For each multi-track the vocals and all the other music instrument sources are used collectively. The accompaniment source is computed by adding the bass, drums, and other music instrument sources. Then, each sampled multi-track is down-mixed to a single channel and is partitioned into overlapping segments of $N=44100$ samples. The overlap is 22050 samples. The segments for each source are independently and randomly shuffled. Then, the singing voice signal segments are corrupted using the shuffled segments of the accompaniment source. For the corruption by additive Gaussian noise, the standard deviation of the noise remains constant and is independent from the amplitude of the singing voice signal. For optimizing the parameters of the representation learning method, with respect to the minimization of Eq.~\eqref{eq:tot-loss-TV} or Eq.~\eqref{eq:tot-loss-SK}, the adam algorithm~\cite{adam} is used. To compute the Sinkhorn distance(s), Eq.~\eqref{eq:pairwise-distance} is applied to each $\mathbf{A}_{\text{m}}$ contained within the batch, and the average distance is computed within each batch.

\subsection{Evaluation}\label{ch5:sec:exp_proc_testing}
For evaluating the usefulness of the representation that is learned by the proposed method,
the rest of the $50$ tracks from the MUSDB18 data-set are used. Each track is down-mixed and partitioned into non-overlapping segments of $N=44100$ samples. Shuffling and random mixing are not performed at this stage. However, the silent segments in the singing voice tracks are discarded. Discarding silent singing voice segments is based on:
 \[ l_{\mathrm{x}_{\text{v}}} = 10 \text{log}_{10} (|| \mathbf{x}_{\text{v}}||_2^2 + \epsilon) \, \, \,
\begin{cases}
      \mathbf{x}_{\text{v}}: \text{active}, & \text{if}\ l_{\mathrm{x}_{\text{v}}}\geq -10 \\
      \mathbf{x}_{\text{v}}: \text{silent}, & \text{otherwise,}
\end{cases}
\]
where $l_{\mathrm{x}_{\text{v}}}$ is the thresholding value for discarding a segment. The thresholding value is empirically chosen by finding the minimum value, in the training sub-set of MUSDB18 for all segments, that can be used to preserve all active singing voice segments.

The representation is evaluated with respect to the three following criteria: i) reconstruction error of the proposed method to encode and decode the clean singing voice signal using the previously described methodology, ii) reconstruction error of the separated singing voice signal by binary masking, and iii) additivity of the representation. The first and second criteria are  objectively measured with respect to the clean singing voice signal $\mathbf{x}_{\text{v}}$ using the scale invariant signal-to-distortion ratio (SI-SDR)~\cite{si_sdr}. The SI-SDR, expressed in dB, is computed for each segment as
\begin{align}\label{eq:si-sdr}
    \text{SI-SDR}(\mathbf{x}_{\text{v}}, \hat{\mathbf{x}}_{\text{v}}) &= 10\,\text{log}_{10} \Big(\frac{||\alpha \mathbf{x}_{\text{v}}||_{2}^2}{||\alpha\mathbf{x}_{\text{v}}-\hat{\mathbf{x}}_{\text{v}}||_{2}^2}\Big) \text{, for } \nonumber \\ \alpha &= \frac{\tilde{\mathbf{x}}_{\text{v}}^{T}\mathbf{x}_{\text{v}}}{||\mathbf{x}_{\text{v}}||_{2}^{2}}\text{.}
\end{align}
Higher SI-SDR values indicate better reconstruction or separation performance. It should be noted that the first criterion is used only to evaluate the reconstruction capabilities of the proposed re-parameterization scheme and not the learning capabilities of the overall method for learning representations. That is because this reconstruction criterion does not support the claim of the proposed method to be unsupervised, since the reconstruction of the singing voice has been used as an optimization objective; yet it serves as an informative quality indicator for audio signals.

For performing the task of singing voice separation, informed binary masking is used. That is because  
masking is an important operation in audio and music source separation, and has been extensively used by deep learning based approaches and also representation learning~\cite{tzinistwostep}. The focus is given on informed separation, i.e., masks are computed by an oracle method using the information for all the mixture's sources available in the data-set. This is done in order to estimate the least-upper-bound performance of singing voice separation, for a learned representation. This alleviates the biases on the prior information that music source separation approaches have. Examples of biases include the source's structure and the existing neural architectures that are engineered for the representations computed using the STFT. Finally, binary masking is used because it is an indicator of how disjoint (less overlap) two sources are, given a representation\footnote{For the detailed connection between disjointness and binary masking see the Appendix.}. More specifically, the oracle binary masking is computed by encoding three signals. The first signal is the mixture $\mathbf{x}_{\text{m}}$, the second signal is the accompaniment source $\mathbf{x}_{\text{ac}}$, and the singing voice signal $\mathbf{x}_{\text{v}}$. The representations $\mathbf{A}_{\text{m}}$, $\mathbf{A}_{\text{ac}}$, and $\mathbf{A}_{\text{v}}$ of the signals $\mathbf{x}_{\text{m}}$, $\mathbf{x}_{\text{ac}}$, and  $\mathbf{x}_{\text{v}}$, respectively, are computed using the trained encoder $E(\cdot)$. The mask $\mathbf{G}_{\text{v}} \in \mathbb{R}^{C \times T}$ is computed as
\[\mathbf{G}_{\text{v}} = g(\mathbf{A}_{\text{v}} \, \oslash \, \mathbf{A}_{\text{ac}}) \text{ , } \]
where $g(\cdot)$ is defined as
\[g(\mathrm{x}) = 
    \begin{cases}
      1, & \text{if}\ \mathrm{x}\geq0.5 \\
      0, & \text{otherwise}
    \end{cases}
\text{ . }\]
The approximation of the singing voice time-domain signal $\hat{\mathbf{x}}_{\text{v}}$ using the decoder $D(\cdot)$ and by means of binary masking is computed as
\[\hat{\mathbf{x}}_{\text{v}} = D(\mathbf{A}_{\text{m}} \odot \mathbf{G}_{\text{v}})
\text{ . }\]
%where ``$\odot$'' is the element-wise (Hadamard) product.

The additivity of the sources is computed using the following objective metric
\begin{equation}\label{eq:adt}
    \mathcal{A}(\mathbf{x}_{\text{m}}, {\mathbf{x}}_{\text{v}}, {\mathbf{x}}_{\text{ac}}) = 1 - \frac{||E(\mathbf{x}_{\text{m}}) - E(\mathbf{x}_{\text{v}}) - E(\mathbf{x}_{\text{ac}})||_1}{||E(\mathbf{x}_{\text{m}})||_1  + \varepsilon} \text{ , } 
\end{equation}
where $||\cdot||_1$ is the $L_1$ matrix norm, $\varepsilon = 1e-24$ is a small term for ensuring numerical stability, and  $\mathbf{x}_{\text{ac}}$ is the time-domain signal of the accompaniment music source that is computed by mixing the multi-tracks available in the testing subset. High values of $\mathcal{A}(\cdot)$ indicate that the representation of the mixture signal consists of additive sources (higher $\mathcal{A}(\cdot)$ is better). The attribute of additivity is important for the computation of optimal separation masks~\cite{liutkus_alpha}, and in the unsupervised separation of music sources~\cite{Smaragdis06aprobabilistic, huang_rpca}.

\subsection{Assessing Design Choices}
Using the procedures that are described in Section~\ref{ch5:sec:exp_proc_training} and Section~\ref{ch5:sec:exp_proc_testing}, two additional experiments are conducted. For both experiments every model is optimized three times using different initial random seeds. For the first experiment, the modulated cosine functions (\texttt{mod-cos}) are examined for their applicability as synthesis model by measuring the reconstruction performance, after being optimized for the denoising task. It should be noted that for the first experiment the corruption process with the randomly shuffled segments of the accompaniment signal is not considered. Furthermore, an early stopping mechanism is used to terminate the training procedure if the model under examination has stopped decreasing the reconstruction objective (neg-SNR), expressed in Eq.~\eqref{eq:neg-snr}, on average with respect to the batches in the previous iteration. For comparison, various modifications to the presented method for representation learning and decoding strategies from related literature are considered in this experiment. Specifically, the squaring of the normalized frequencies $f_c$ is examined, among other decoding strategies such as non-modulated cosine functions (\texttt{cos}), and common one-dimensional convolutional networks (\texttt{conv}) with and without the tanh non-linearity applied at the last stage of the decoding process. In addition to this, Sinc-Net~\cite{sinc_net} (\texttt{sinc}) Sinc-Net is examined as the first encoding stage as proposed in~\cite{sinc_net}. For this experiment, $C$ is adapted for each model so that the same number of parameters is used by the models.

The best combination of the decoding and non-linear functions from the first experiment are further investigated in the second experiment. In this experiment the following values for the number of components $C \in [400, 800, 1600]$ are examined. Furthermore, the effect of the representation objective is examined with respect to the usage of information either from the additive corruption by multi-modal noise or the additive corruption by Gaussian noise, i.e., using either $\mathbf{A}_{\text{v}}$ or $\mathbf{A}_{\text{m}}$. For this experiment, the (an-isotropic) total-variation denoising (Eq.~\eqref{eq:tv_loss}) objective is used, as the Sinkhorn distances are computationally very demanding and significantly slow-down the training procedure. For comparison, the STFT is employed by performing the above described operations of analysis, masking, and synthesis. The STFT uses a hop-size of 256 samples, a window size of 2048 samples, and the hamming windowing function.

\section{Results \& Discussion}\label{ch5:sec:results}
\subsection{Results from Design Choices Evaluation}\label{ch5:sec:results-design-choices}
Table~\ref{app:tab:res-1} demonstrates the median values of SI-SDR expressed in dB (the higher the better) yielded by the first experiment, with additional information regarding the various setups for the encoder $E(\cdot)$ and the decoder $D(\cdot)$, the number of parameters $N_P$ (in millions M), the used number of components $C$, and the employed non-linearities. The results in Table~\ref{app:tab:res-1} highlight three trends. First, the application of the non-linearity to the normalized frequencies $f_c$ results into better reconstruction performance compared to the linear case. The observed improvement is of $\sim5$dB on average across experimental configurations. Secondly, the modulated cosine functions serve as a good differentiable synthesis model for singing voice signals, outperforming simple cosine functions by approximately 8 dB on average, with respect to the two experimental configurations (with and without frequency scaling of the normalized frequency), and by $1.4$ dB the best configuration of convolution based model (\texttt{conv}). Since SI-SDR is invariant to scale modifications of the assessed signal, $1.4$ dB is a significant improvement of signal quality and does not imply a simple matching of the gain that the model based on modulated cosine functions might have exploited. Thirdly, Sinc-Net~\cite{sinc_net} does not bring further improvements to the proposed method.

\begin{table}[!t]
\smallskip
\centering
\caption{Results reflecting the decoding performance, by means of SI-SDR. Bold-faced numbers denote the best performance.}
%\resizebox{0.75\columnwidth}{!}{%
\begin{tabular}{c|c|c|c|c}  
$E(\cdot)/D(\cdot)$ Setup & Non-linearity  & $C$ & SI-SDR & $N_P$    \\ \hline
\multirow{2}{*}{\small{\texttt{conv/cos}}} & N/A & \multirow{2}{*}{952} & 20.83 & \multirow{2}{*}{6.483M} \\
 {} &  $f_{c}^{2}$ & & 22.34 & {} \\ \hline
\multirow{2}{*}{\small{\texttt{conv/conv}}} & N/A & \multirow{2}{*}{800} & 31.25 & \multirow{2}{*}{6.476M} \\
 {} & tanh(decoder) & {} & 30.50 & {} \\ \hline
 \multirow{2}{*}{\small{\texttt{conv/mod-cos}}} & N/A & \multirow{2}{*}{800} & 28.72 & \multirow{2}{*}{6.478M} \\
 {} & $f_{c}^{2}$ & {} & \bf{32.62} & {} \\ \hline
 {\small \texttt{sinc/mod-cos}} & $f_{c}^{2}$ & {952} & 26.82 & 6.487M \\ 
\end{tabular}%}
\label{app:tab:res-1}
\end{table}

Focusing on the separation performance of the obtained representations, Table~\ref{app:tab:res-2} presents the median SI-SDR values of the binary masking separation scenario, for three values for the hyper-parameter $C$ and two strategies for computing the representation objective. These strategies consider two different signal representations that are either the corrupted by Gaussian noise $\mathbf{A}_{\text{v}}$ or the synthetic mixtures using the accompaniment signals $\mathbf{A}_{\text{m}}$. The obtained results are compared to the common convolutional encoder/decoder setup used in Table~\ref{app:tab:res-1} and the STFT that has perfect reconstruction properties and masking techniques work very well in practice~\cite{rafii18}. The results of Table~\ref{app:tab:res-2} mainly underline two experimental findings. The main finding is that using the representation objective and information from the realistic corruption process $\mathbf{A}_{\text{m}}$, it can be used to improve the reconstruction of the masked mixture signals without additional supervision, as previous studies suggest~\cite{tzinistwostep}. This claim is supported by the observed improvement of $\sim 2$ dB, on average across models of various components $C$, when the synthetic mixtures are used for the unsupervised representation objective. Furthermore, the proposed re-parameterization scheme improves by approximately 1.6 dB the separation performance compared to typical convolutional networks. Nonetheless, there is much room for improvements in order to obtain the quality of the STFT/iSTFT approach that outperforms the best masked approximation of the proposed method by $2.12$ dB.
\begin{table}[!t]
\smallskip
\centering
\caption{SI-SDR for informed separation by binary masking (BM). Bold-faced numbers denote the best performance.}
%\resizebox{0.85\columnwidth}{!}{%
\begin{tabular}{c|c|c|c|c|c}  
$E(\cdot)/D(\cdot)$ Setup & $C$ & ${\mathcal{L}_{\text{TV}}}(\nicefrac{{}^{\mathbf{A}_\text{m}}}{{}_{\mathbf{A}_\text{v}}})$ & SI-SDR & SI-SDR-BM & $N_P$    \\ \hline
\multirow{6}{*}{\small{\texttt{conv/mod-cos}}} & \multirow{2}{*}{400} & $\mathbf{A}_\text{v}$ & 30.46 & 3.66 & \multirow{2}{*}{2.439M} \\
{} & {} & $\mathbf{A}_\text{m}$ & 30.73 & 5.93 & {} \\\cline{2-6}
{} & \multirow{2}{*}{800} & $\mathbf{A}_\text{v}$ & \bf{32.28} & 4.39 & \multirow{2}{*}{6.478M} \\
{} & {} & $\mathbf{A}_\text{m}$ & 32.11 & 6.28 & {} \\\cline{2-6}
{} & \multirow{2}{*}{1600} & $\mathbf{A}_\text{v}$ & 31.94 & 4.68 & \multirow{2}{*}{19.356M} \\ 
{} & {} & $\mathbf{A}_\text{m}$ & 31.54 & 6.68 & {} \\ 
\hline
\multirow{2}{*}{\small\texttt{conv/conv}} & \multirow{2}{*}{800} & $\mathbf{A}_\text{v}$ & 31.25 & 2.89 & \multirow{2}{*}{6.476M} \\
{} & {} & $\mathbf{A}_\text{m}$ & 31.13 & 4.95 & {} \\
\hline
{\small\texttt{STFT/iSTFT}} & 1025 & N/A & N/A & \bf{8.80} &  N/A %8.388M
\end{tabular}%}
\label{app:tab:res-2}
\end{table}

\subsection{Representation Learning Results}
Table~\ref{tab:res-1} presents the average and standard deviation values of the additivity measure $\mathcal{A}(\cdot)$, the SI-SDR for the reconstruction and the separation objective performance in dB, and the values of the hyper-parameters $\omega$ and $\lambda$ used to compute the two representation objectives. The results in Table~\ref{tab:res-1} are
discussed according to 
the SI-SDR value (higher is better), because SI-SDR assesses the reconstruction and separation performance. 
\begin{table}[!th]
\smallskip
\centering
\caption{Results from objectively evaluating the learned representations. Values in boldface denote the best obtained performance.}
\begin{tabular}{c|c|c|c|c|c}  
Objective & $\omega$  & $\lambda$ & SI-SDR (dB) & SI-SDR-BM (dB) & $\mathcal{A}(\cdot)$ \\
\midrule
\multirow{5}{*}{${L}_{A}$} & 0.5 & N/A & 31.49 ($\pm 2.98$) & 4.43 ($\pm 4.98$) & 0.76 ($\pm 0.10$) \\
{} & 1.0 & N/A & 31.39 ($\pm 3.16$) & 4.66 ($\pm 4.92$) & 0.76 ($\pm 0.10$) \\
{} & 1.5 & N/A & 31.01 ($\pm 3.13$) & 4.97 ($\pm 4.93$) & 0.75 ($\pm 0.10$) \\
{} & 2.0 & N/A & 30.96 ($\pm 2.98$) & 4.65 ($\pm 4.90$) & 0.76 ($\pm 0.10$) \\
{} & 4.0 & N/A & 31.40  ($\pm 2.83$) & 5.06 ($\pm 4.97$) & 0.76 ($\pm 0.10$) \\
\hline
\multirow{5}{*}{${L}_{B}$} & 1.0 & 0.1 & 31.28($\pm 2.98$) & 5.40($\pm 5.31$) & 0.76($\pm 0.09$) \\
{} & 1.0 & 0.5 & $\mathbf{31.61}(\pm 3.38)$ & $\mathbf{5.63 (\pm 5.29)}$ & 0.77($\pm 0.09$)  \\
{} & 1.0 & 1.0 & 31.29($\pm 3.25$) & 4.33($\pm 5.28$) & 0.86($\pm 0.08$) \\
%{} & 1.0 & 1.3 & 30.99($\pm 3.53$) & 0.17($\pm 6.42$) & 0.89($\pm 0.08$) \\
{} & 1.0 & 1.5 & 29.98($\pm 3.48$) & 0.06 ($\pm 6.43$) & $\mathbf{0.89(\pm 0.08)}$ \\
{} & 1.0 & 2.0 & 31.13($\pm 3.66$) & -0.02($\pm 6.44$) & $\mathbf{0.89(\pm 0.08)}$
\end{tabular}
\label{tab:res-1}
\end{table}

There are two observable trends in Table~\ref{tab:res-1}. The first trend is that when using ${L}_{B}$, small values of $\lambda$
marginally improve the SI-SDR, compared to the best SI-SDR when using ${L}_{A}$ (i.e., $\omega=0.5$ and SI-SDR=31.49). Specifically, when using ${L}_{B}$ as the representation objective and for $\lambda = 0.5$, the SI-SDR and SI-SDR-BM are improved by $0.12$ dB and 1.20 dB, respectively, compared to the  case of using ${L}_{A}$ and $\omega=0.5$. Additionally, with the same $\lambda=0.5$ for ${L}_{B}$, an improvement of $0.57$ dB SI-SDR-BM can be observed, compared to the best obtained SI-SDR-BM using ${L}_{A}$ with $\omega=4.0$. This trend shows that when using the Sinkhorn distances as an objective (i.e., ${L}_{B}$) with a small entropic regularization weight, i.e., small values of $\lambda$, there is a marginal improvement of the reconstruction performance for the singing voice (measured with SI-SDR-BM), but also the learned representations yield better results for singing voice separation (measured with SI-SDR).

The second trend observed in Table~\ref{tab:res-1} is that when using ${L}_{B}$ and $\lambda>1$, specifically for $\lambda\in[1.5, 2.0]$, the SI-SDR for binary masking drops by more than $5$ dB, compared to ${L}_{B}$ with $\lambda=0.5$. This indicates that the separation by binary masking fails, suggesting that the singing voice and accompaniment are completely overlapping in the representation of the mixture $\mathbf{A}_{\text{m}}$. That is expected since entropy expresses the uncertainty about the representation of the mixture signal. This means that during training, all the components of the representation are equally probable to be active when the mixture signal is encoded. Interestingly enough, that uncertainty in the encoding process comes with the observed effect that the sources become additive in the learned representation.

To further investigate the effect of entropic regularization with respect to the additivity metric, the impact of the weight $\omega$ on ${L}_{B}$ is examined. To that aim, the best $\lambda=1.5$ from Table~\ref{tab:res-1} is chosen as a fixed hyper-parameter and $\omega$ is varied. The corresponding results are given in Table~\ref{tab:res-2} and are compared to the magnitude representation computed using the STFT, that is the most commonly employed representation for music source separation. The results from Table~\ref{tab:res-2} suggest that by increasing the weight $\omega$ that affects the strength of the representation objective in the learning signal, the learned mixture representations consist of two almost additive representations, i.e., the singing voice and the accompaniment representations. This is observed for $\omega=4.0$. Furthermore, nearly all representations computed using the Sinkhorn distances and the entropic regularization, outperform the magnitude of the STFT with respect to the objective measure of additivity in an unsupervised fashion, i.e., additivity was not explicitly enforced using an optimization objective.
\begin{table}[!th]
\smallskip
\centering
\caption{Objective evaluation of the additivity of the learned representations.}
\label{tab:res-2}
%\resizebox{0.6\columnwidth}{!}{%
\begin{tabular}{c|c|c|c}  
% \toprule
Objective & $\omega$  & $\lambda$ & $\mathcal{A}(\cdot)$ \\
\hline
\multirow{4}{*}{${L}_{B}$} & 1.0 & 1.5 & 0.89 ($\pm 0.08$) \\
{} & 1.5 & 1.5 & 0.90 ($\pm 0.07$) \\
{} & 2.0 & 1.5 & 0.92 ($\pm 0.07$) \\
{} & 4.0 & 1.5 & $\mathbf{0.93 (\pm 0.06)}$ \\
% \bottomrule
\hline
STFT & N/A & N/A & 0.86 ($\pm 0.06$)
\end{tabular}%
%}
\end{table}

\begin{figure}[!ht]
\centering
\begin{minipage}{0.99\columnwidth}
\centering
 \includegraphics[width=0.33\columnwidth]{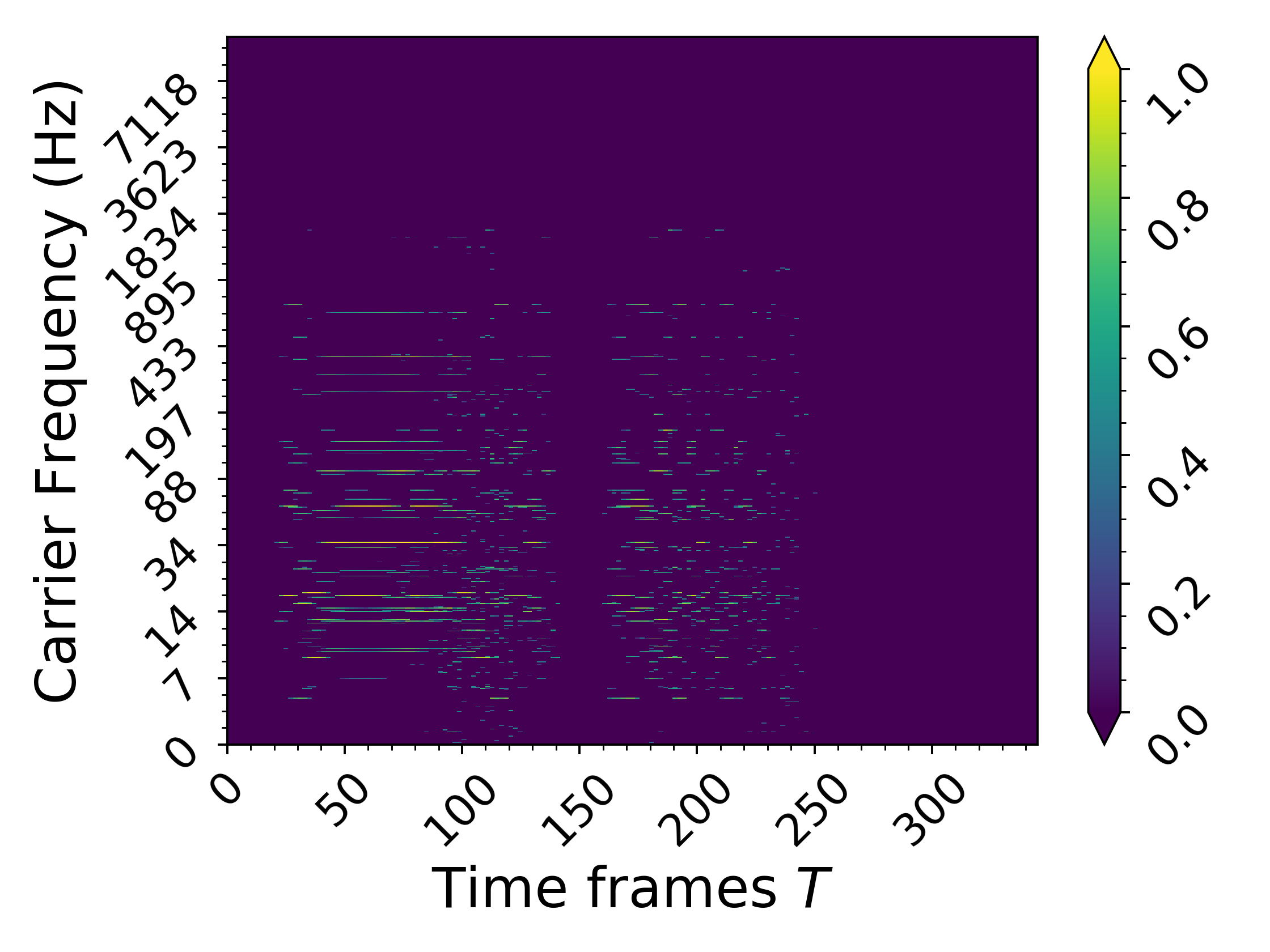}%
 \includegraphics[width=0.33\columnwidth]{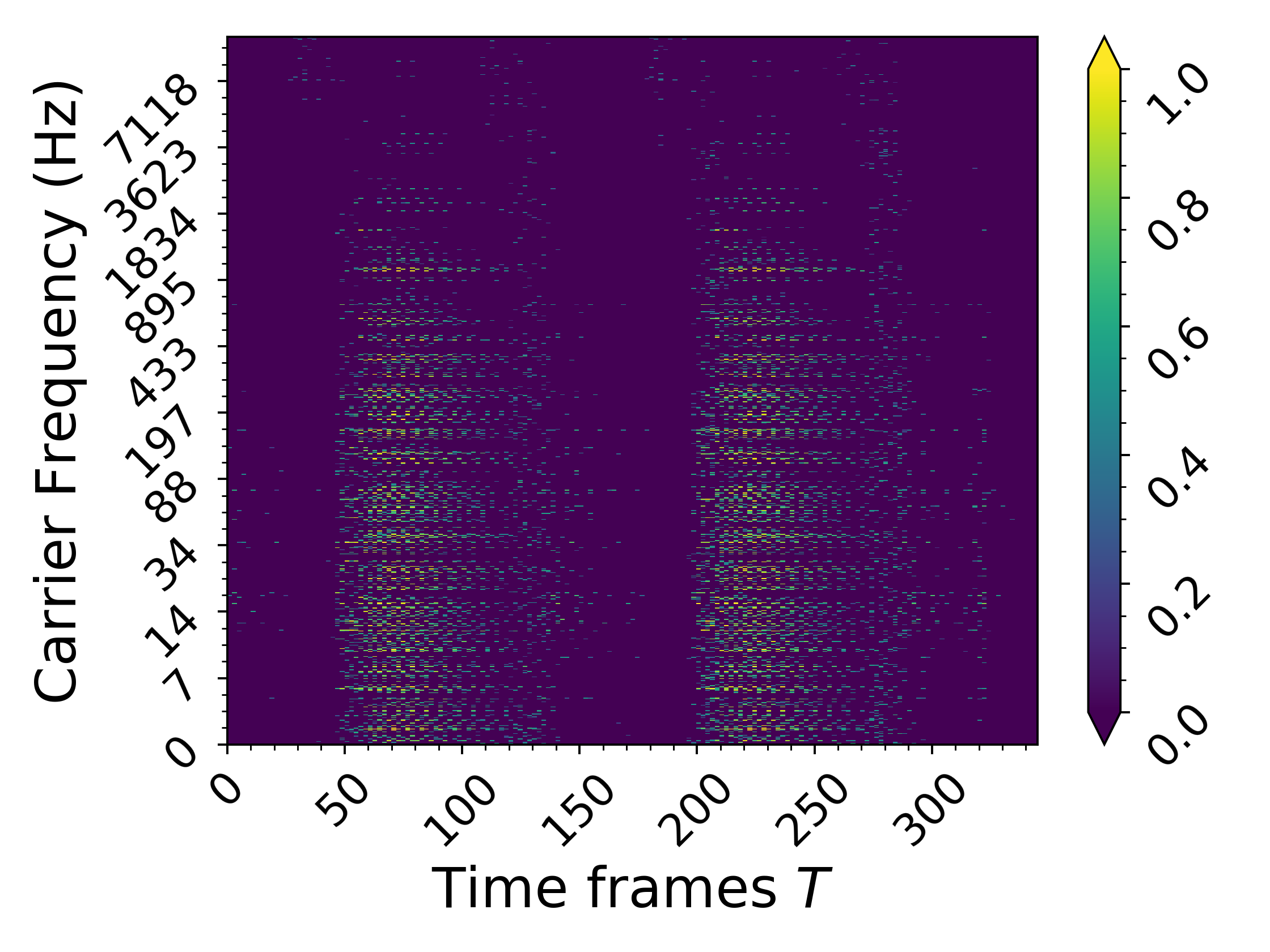}%
 \includegraphics[width=0.33\columnwidth]{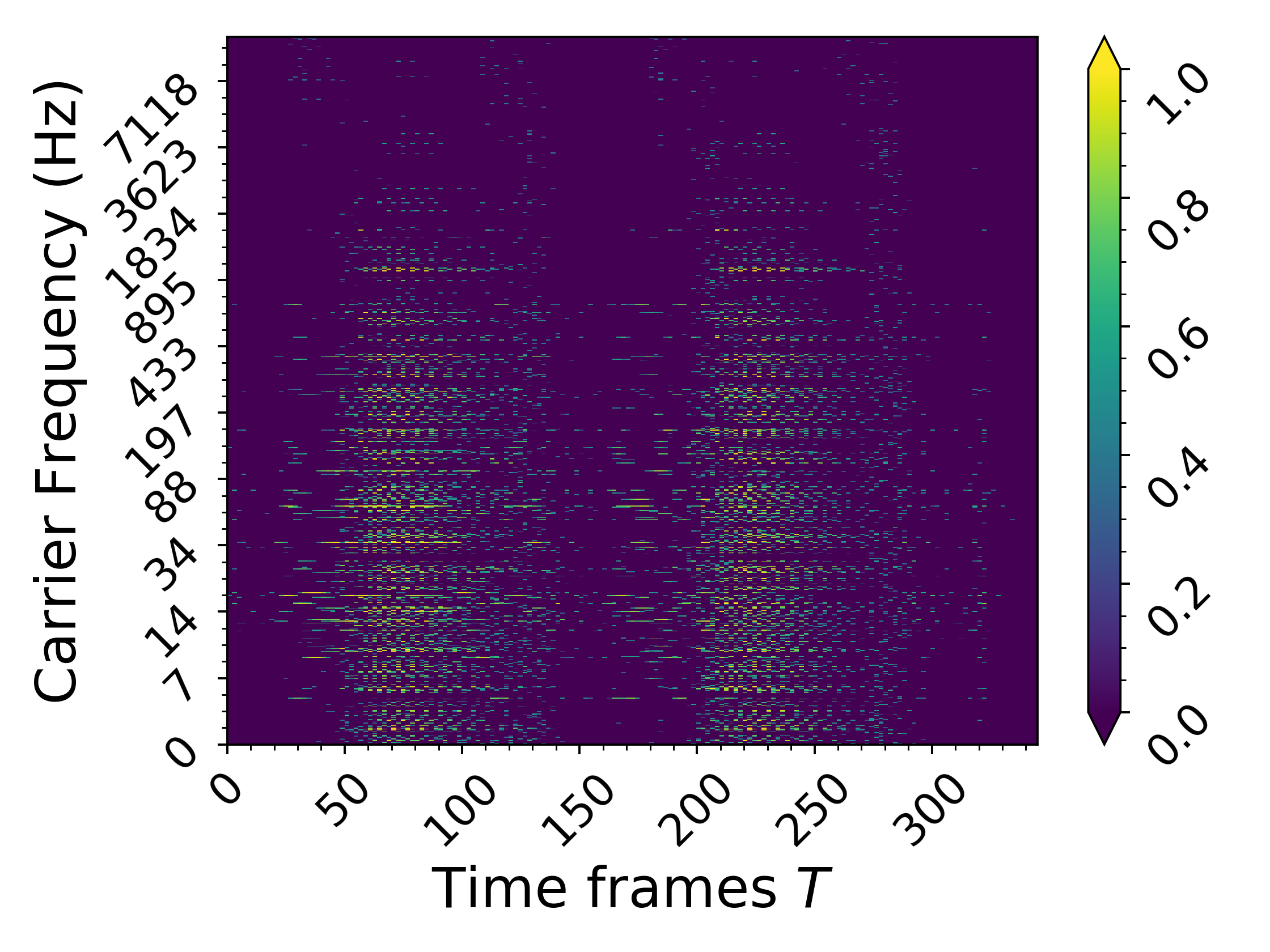}%
 \centering
 {\small \subcaption{Learned representations for the singing voice (top-left), the accompaniment (top-right), and the mixture (bottom-middle) signals using the $E(\cdot)$ optimized with ${L}_{A}$ for $\mathcal{L}_{\text{TV}}(\cdot) \text{ with } \omega=4.0$}}%
\end{minipage}\\
\begin{minipage}{0.99\columnwidth}
\includegraphics[width=0.33\columnwidth]{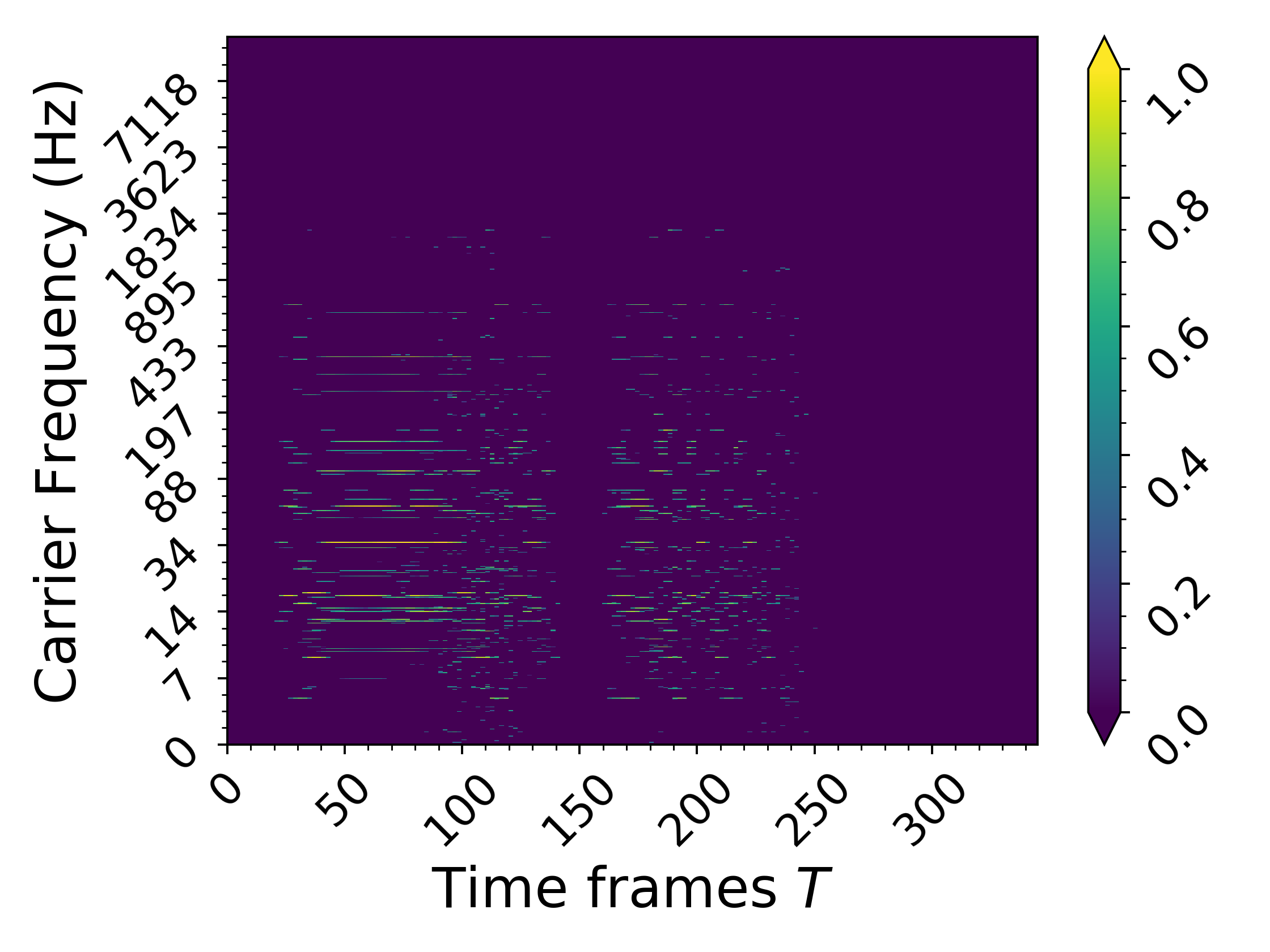}%
\includegraphics[width=0.33\columnwidth]{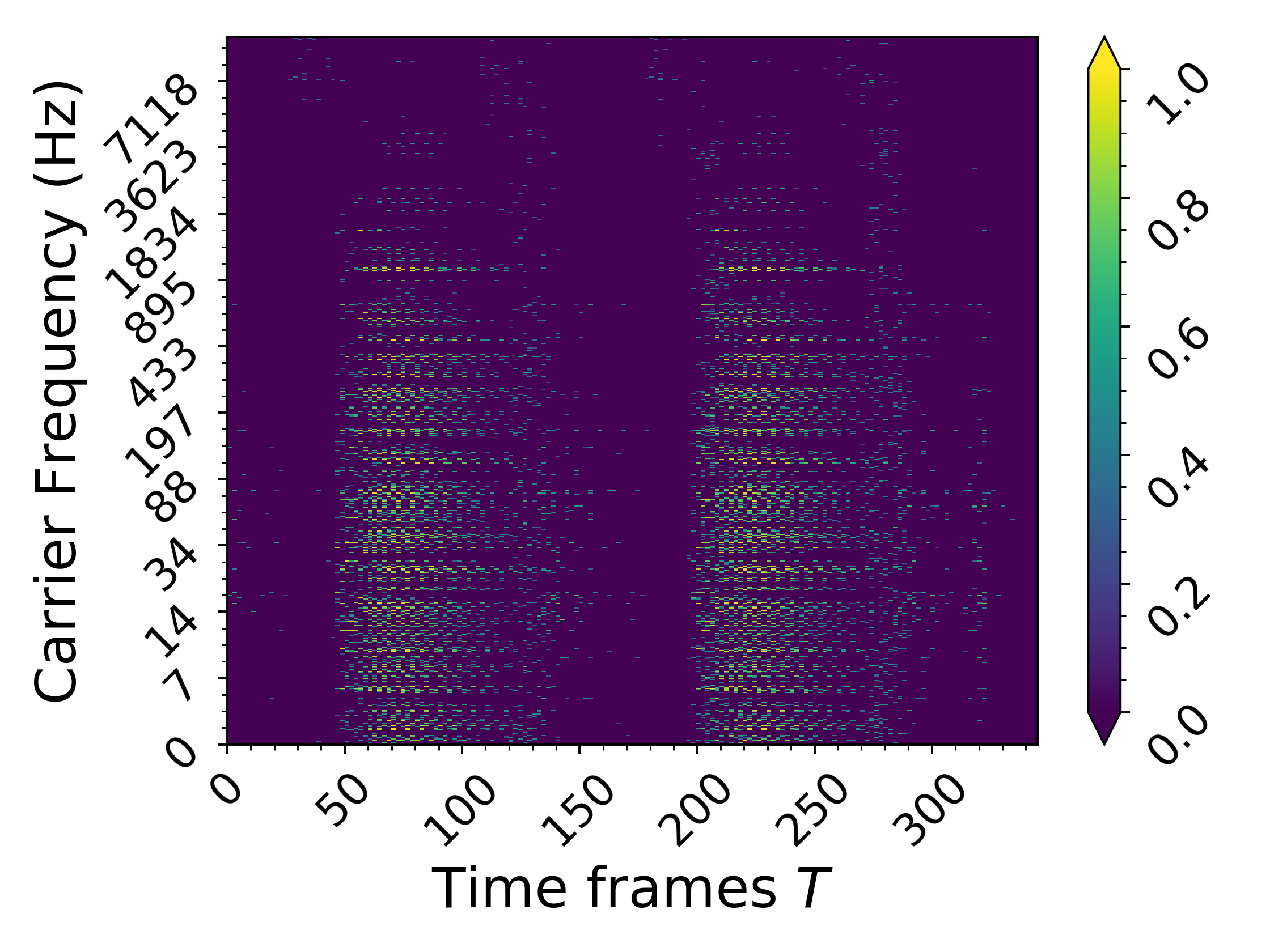}%
 \centering \includegraphics[width=0.33\columnwidth]{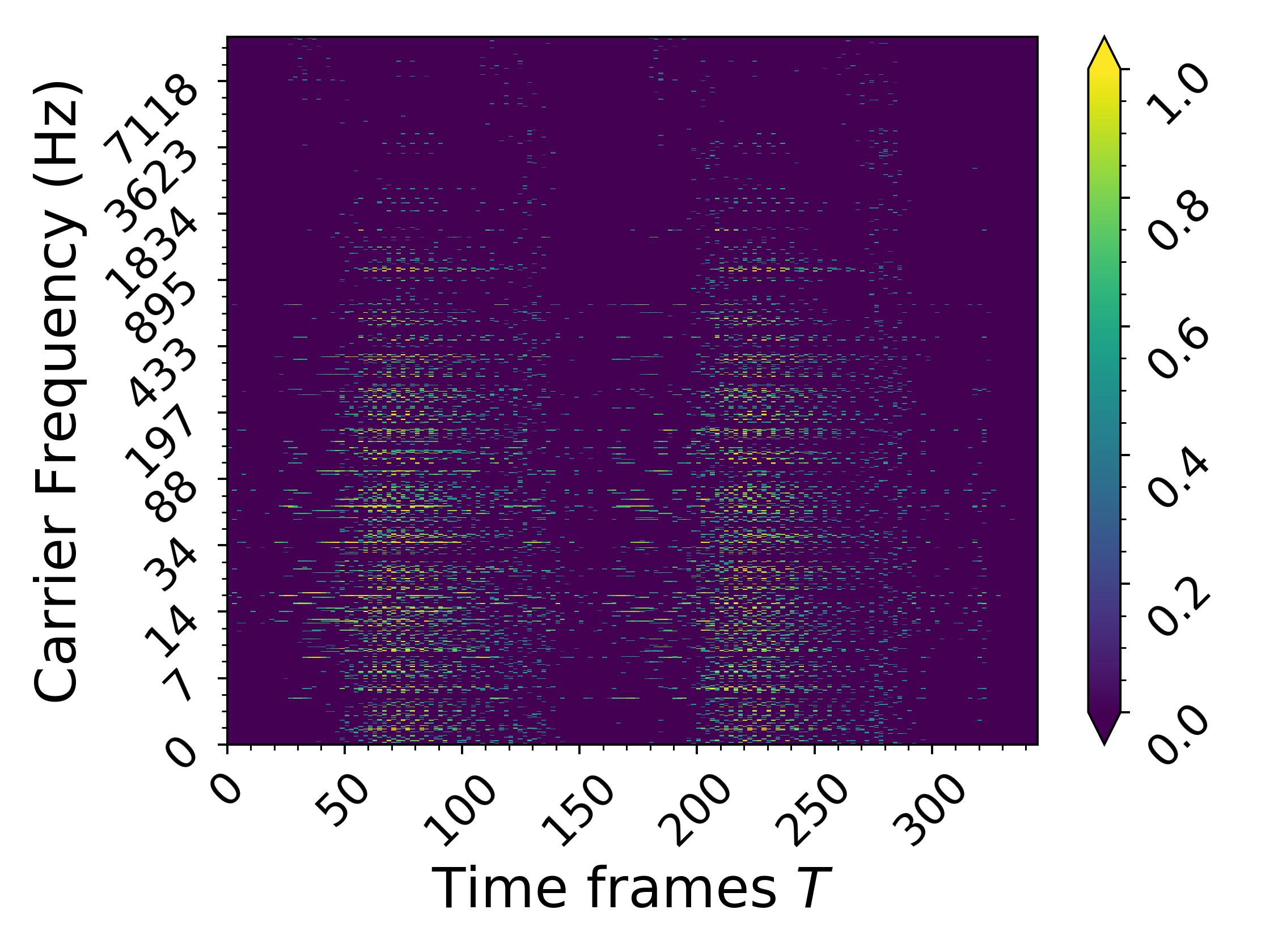}
{\small \subcaption{Learned representations for the singing voice (top-left), the accompaniment (top-right), and the mixture (bottom-middle) signals using the $E(\cdot)$ optimized with ${L}_{B}$ for $\mathcal{L}_{\text{SK}}(\cdot) \text{ with } \omega=1.0, \text{ and } \lambda=0.5$}}
\end{minipage}
\centering
\begin{minipage}{0.99\columnwidth}
\includegraphics[width=0.33\columnwidth]{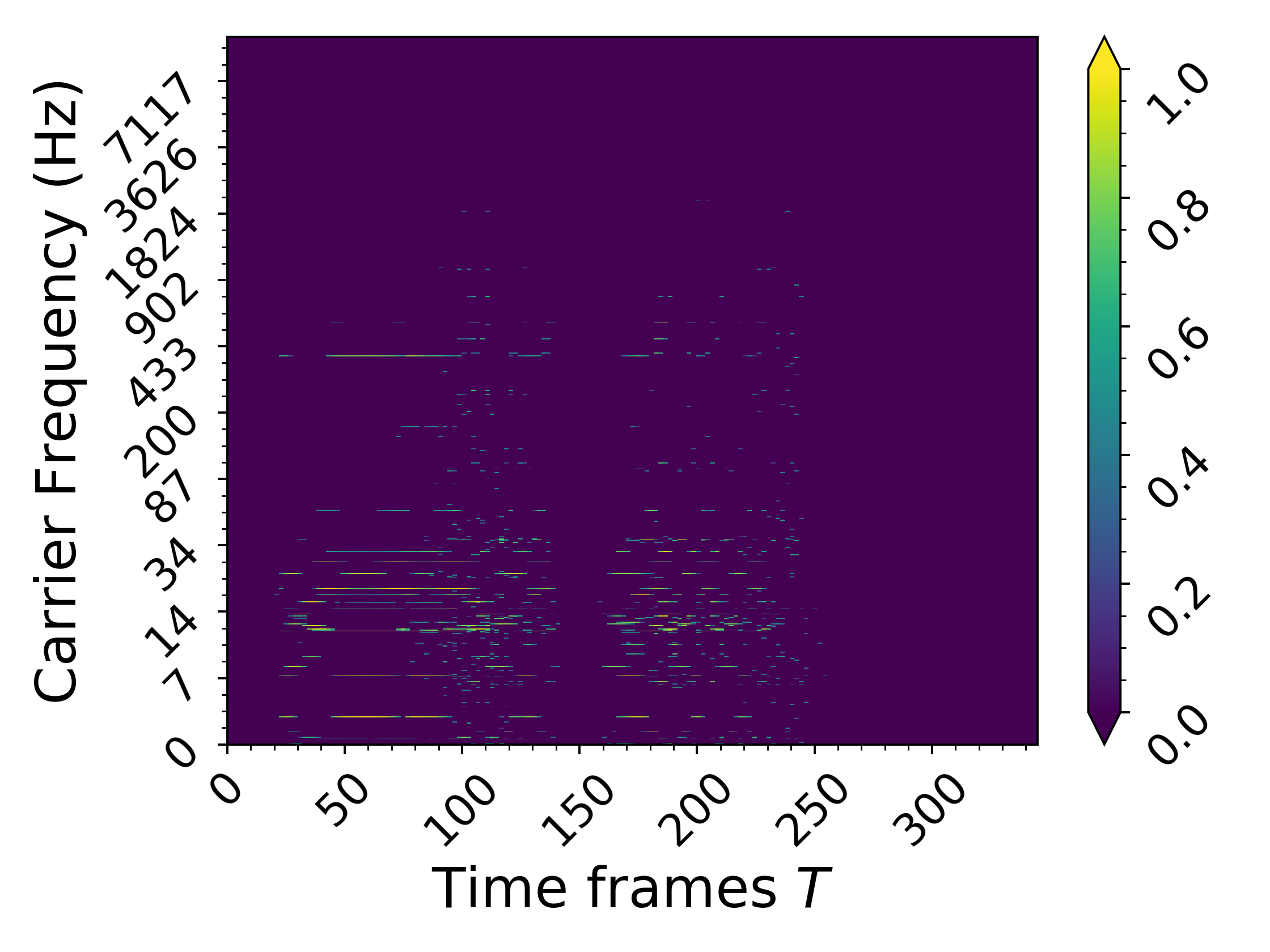}%
\includegraphics[width=0.33\columnwidth]{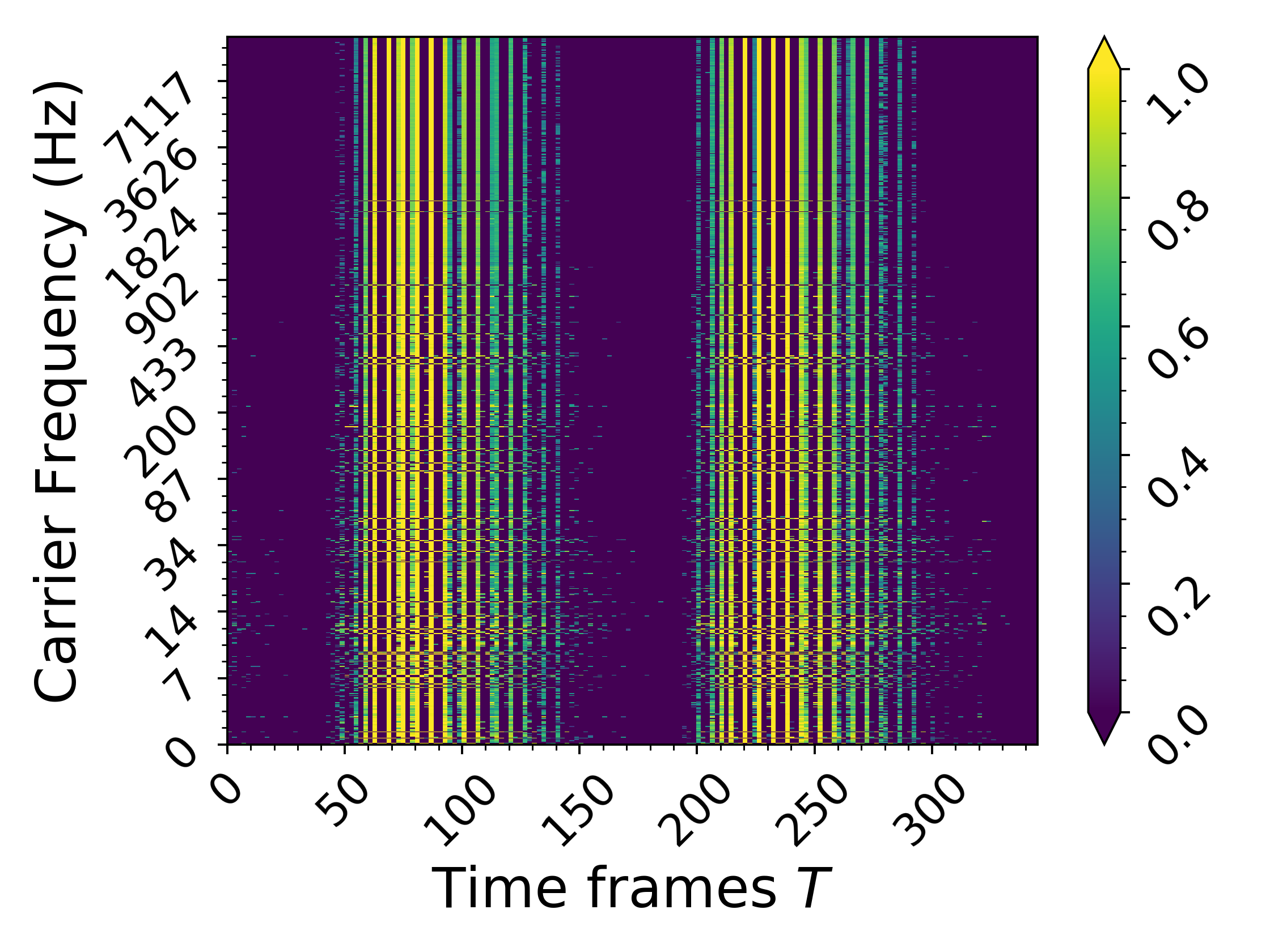}%
\includegraphics[width=0.33\columnwidth]{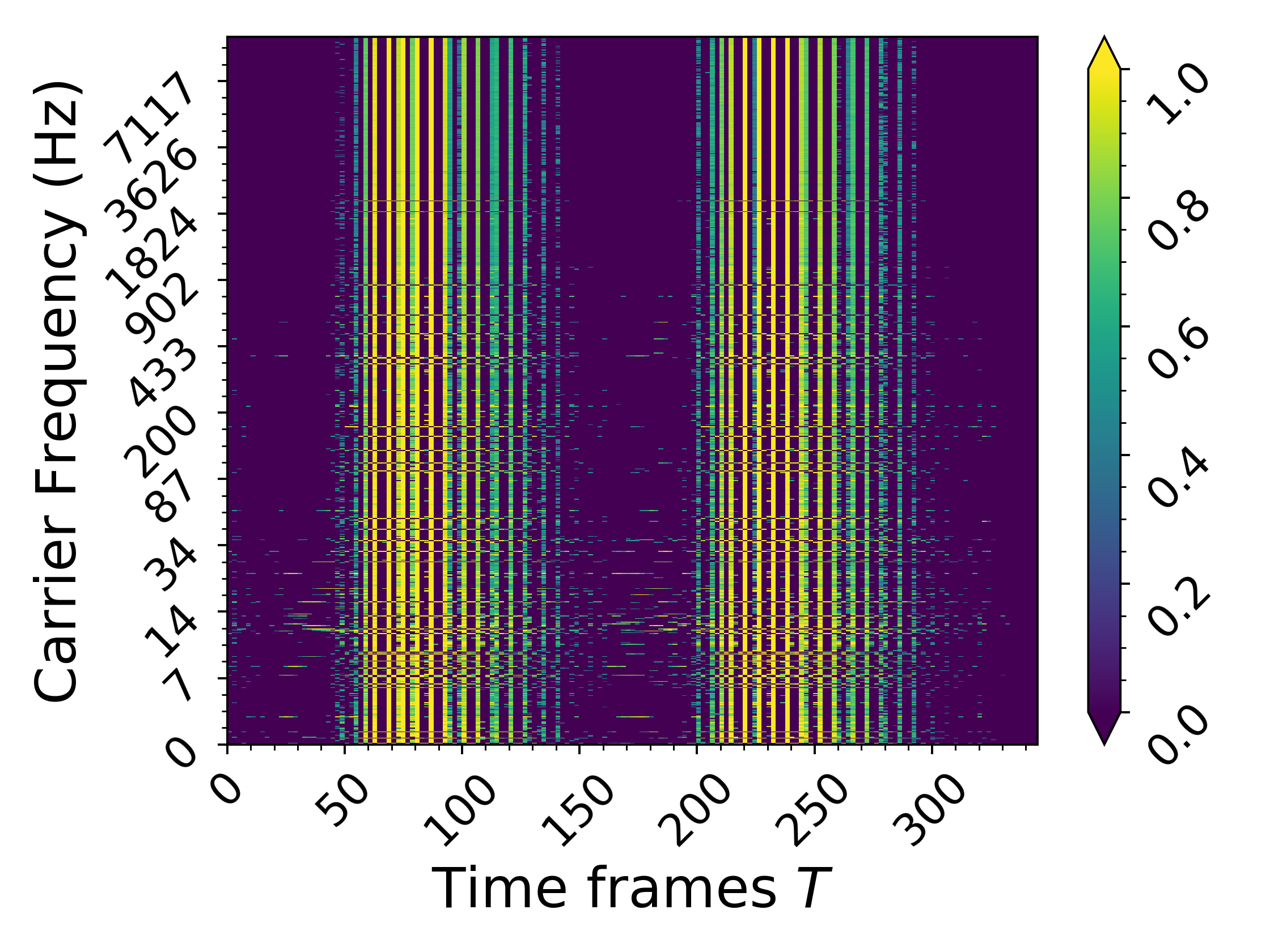}%
{\small \subcaption{Learned representations for the mixture (left), the singing voice (middle), and the accompaniment (right) signals using the $E(\cdot)$ optimized with ${L}_{B}$ for $\mathcal{L}_{\text{SK}}(\cdot) \text{ with } \omega=4.0,\text{ and } \lambda=1.5$}}%
\end{minipage}
\caption{An illustration of the learned representations of a multi-track segment, by three encoders $E(\cdot)$ optimized using various hyper-parameters for ${L}_{A}$ and ${L}_{B}$.}
\label{fig:spectrograms}
\end{figure}

To qualitatively assess the representations for the extreme case observed in Table \ref{tab:res-2}, Figure~\ref{fig:spectrograms} illustrates the learned representations for the mixture, singing voice, and the accompaniment signal using either ${L}_{A}$ or ${L}_{B}$. The signals were acquired from a single multi-track segment contained in the testing sub-set of MUSDB18. For ${L}_{B}$ the focus is given on two extreme cases of separation and additivity performance, that was observed in Table~\ref{tab:res-1} and Table~\ref{tab:res-2}. In particular, Figure~\ref{fig:spectrograms} illustrates the representations obtained
for entropy values $\lambda=1.5$ and for $\lambda=0.5$, that resulted in the best performance of additivity and masking, respectively. For comparison, the learned representations using ${L}_{A}$ are displayed for $\omega=4.0$, which yields the best separation performance according to Table~\ref{tab:res-1}.

In Figure~\ref{fig:spectrograms}(a) it can be clearly observed that the usage of ${L}_{A}$ (employing the total-variation denoising loss) leads to smooth representations. However, qualitatively the representation of the mixture and of the sources seem somewhat blurry, without distinct structure. Consequently, representations learned using ${L}_{A}$ might impose difficulties for source separation methods that aim at capturing the structure of the target music source(s). On the other hand, the employment of ${L}_{B}$ with the Sinkhorn distances and for $\lambda = 0.5$, leads to learned representations that at least for the singing voice signal a prominent structure of horizontal activity is observed. The interesting part comes when the entropy regularization weight is increased to $\lambda = 1.5$. Values of entropic regularization higher than 0.5, enable the learning of representations that for particular sources such as the accompaniment, exhibit distinct structure, i.e., vertical activity (activity with respect to $C$). Furthermore, the representation of the singing voice is characterized by horizontal activity, i.e., a few components $C$ are active and smoothly vary in time. The observed representation structures could be useful for unsupervised separation or audio in-painting methods, such as the deep audio prior~\cite{deep_audio_priors} and the harmonic convolution(s) model~\cite{audio_priors_harm_convs}.

\subsubsection*{On Representation Interpretability}\index{interpretability of the representation}
An important attribute of the learned representation(s), by using the proposed method, is the interpretability, i.e., the learned representations convey information about functions whose parameters have physical quantities such as frequency for example. This can be seen by inspecting closer Figure~\ref{fig:spectrograms}, where each component $C$, i.e., each row of the spectrogram-like illustration, has a carrier frequency that is expressed in Hz. This rationale can be seen as an analogous to common representations such as STFT that has been extensively used in audio signal processing. However, the are two main differences between the proposed method and the STFT.
\begin{figure}[!th]
\centering
\begin{minipage}{0.99\columnwidth}
\centering
\includegraphics[width=0.45\columnwidth,keepaspectratio]{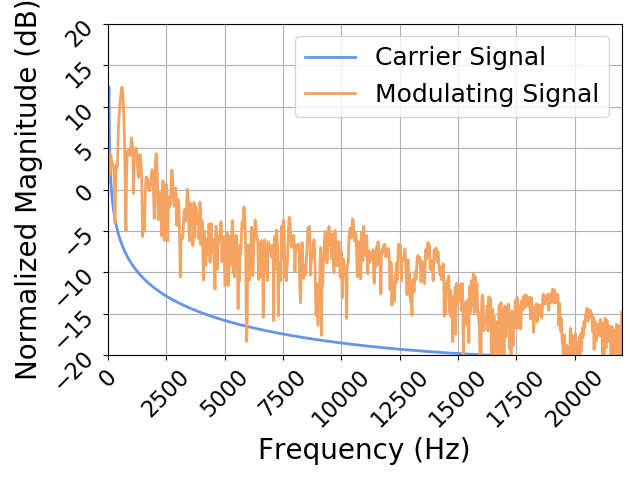}%
 \includegraphics[width=0.45\columnwidth,keepaspectratio]{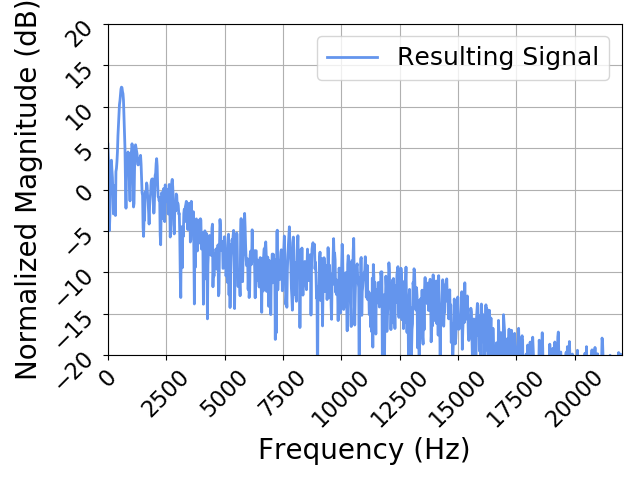}
 {\small \subcaption{Frequency response of a learned basis function, (left) carrier and modulating signal, (right) result of the modulation}}%
 \end{minipage}%
\\
\begin{minipage}{0.99\columnwidth}
\centering
\includegraphics[width=0.45\columnwidth,keepaspectratio]{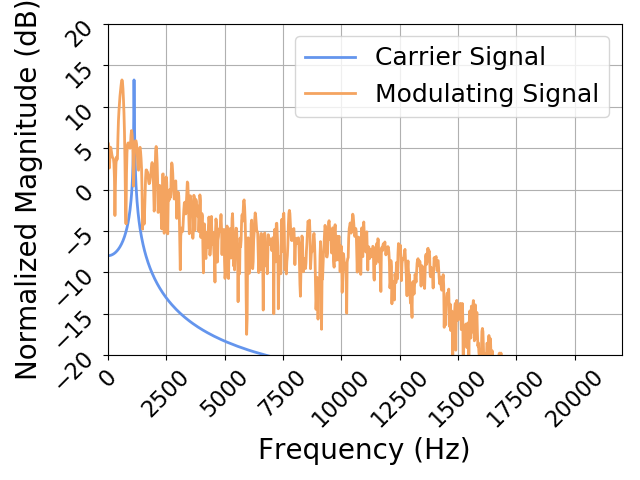}%
 \includegraphics[width=0.45\columnwidth,keepaspectratio]{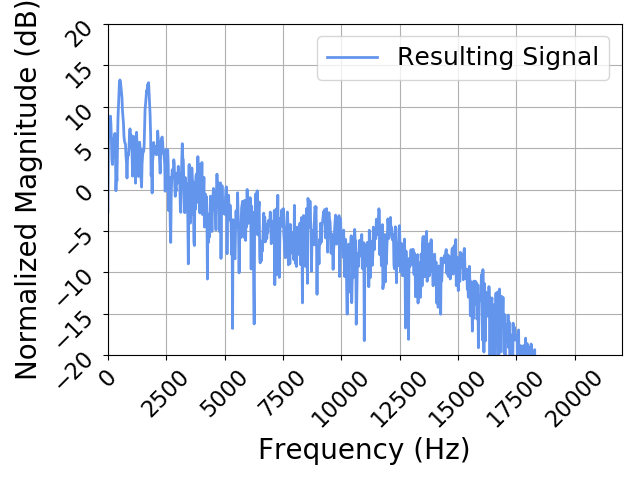}
 \centering
 {\small \subcaption{Frequency response of a learned basis function, (left) carrier and modulating signal, (right) result of the modulation, demonstrating a sinusoidal plus noise structure.}}%
\end{minipage}%
\\
\begin{minipage}{0.99\columnwidth}
\centering
\includegraphics[width=0.45\columnwidth,keepaspectratio]{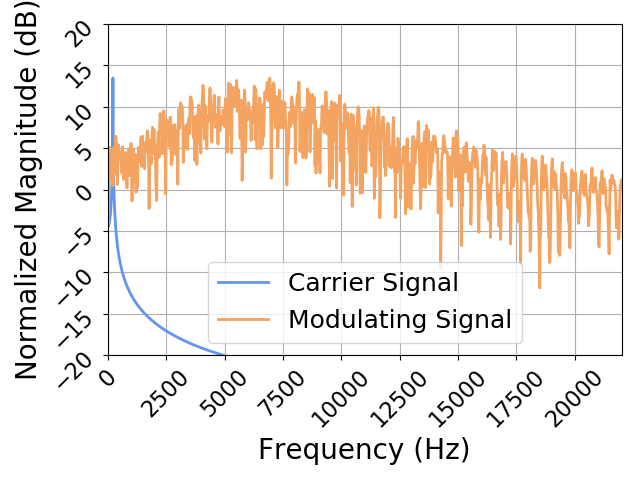}%
 \includegraphics[width=0.45\columnwidth,keepaspectratio]{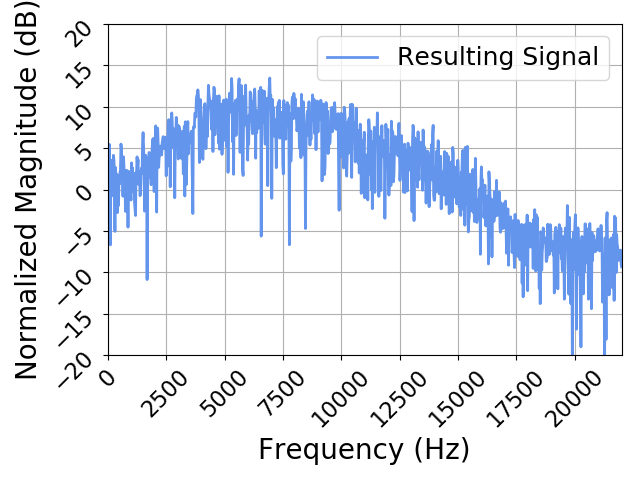}
 \centering
 {\small \subcaption{Frequency response of a learned basis function, (left) carrier and modulating signal, (right) result of the modulation, demonstrating a harmonic plus noise structure.}}%
\end{minipage}%
\caption{The frequency response of three frequently used basis functions that are learned by the proposed method using ${L}_{B}$ with $\lambda=1.5$ and $\omega=4.0$. The frequency response is computed using the discrete Fourier transform, demonstrating a high-frequency comb-like filter.}
 \label{fig:mod-cos-opt-results}
 \end{figure}

The first difference is that the encoding functions of the proposed method are not forced to be symmetric to the decoding functions. This is in contrast to the DFT analysis (encoding) basis functions, employed by the STFT, that are symmetric to the synthesis (decoding) basis functions. This in turn, gives many more degrees of freedom to the encoder of the proposed method, to yield representations that can be optimized with specific objectives. As seen from Figure~\ref{fig:spectrograms}(c) the Sinkhorn distances, with some degree of entropy, allow the computed representations of the accompaniment source to be distinctly structured, something that would not be possible by using symmetric encoding functions. That is because the usefulness of symmetric functions is the perfect reconstruction of a signal after encoding and decoding and not the structure of the output of the encoding, i.e., the representation~\cite{schuller_modulated_pfb, hamidi_pearl_cosines_markovI}. However, this might impose the necessity of devising representation objectives for optimizing the encoder of the proposed method by using domain knowledge from audio and music signal processing.

The second difference is that the decoding functions employed by the proposed method are amplitude-modulated cosine functions as opposed to pure cosine functions that common (audio) transforms have. The main drawback in this case,  is that the modulating signal is directly updated by using back-propagation and it might be hard to interpret after the training procedure. However, the difficulty in interpretation can be tackled by recalling Eq.~\eqref{eq:cos_reparameterization}, in which the signal that is being modulated, i.e., the carrier signal, is a cosine function.
This in turn makes the update rules, based on gradient descent, for the modulating signal to be the linear combination of sinusoidal functions convolved with some noise. That can be verified by evaluating the gradient of the reconstruction error with respect to the modulating signal, that results into a sinusoidal function. Particularly, that function is the convolution of a sinusoid with the derivative of the reconstruction loss with respect to the reconstructed signal. In addition to this, the modulating signal allows an extra degree of freedom in reconstructing signals that cannot be described by pure sinusoidal functions~\cite{serra_sms}, especially when additional representation attributes, such as non-negativity and smoothness, are intended to be learned.
 
To qualitatively assess the information that the modulating functions inherit from the training procedure, Figure~\ref{fig:mod-cos-opt-results} illustrates the frequency response of the carrier and the modulating signal for frequently used components $\mathbf{w}_{c}$ that are in the lower carrier frequency region. The frequency response is obtained by computing the magnitude of the DFT for each corresponding signal. As it can be seen from Figure~\ref{fig:mod-cos-opt-results}, the frequency response of the modulating signal (orange line) consists of a combination of sinusoidal components that have both harmonic structure, considering the position of the observed spectral peaks, but also a stochastic spectral structure. 
The stochastic structure reassembles formants and/or fricatives of the singing voice signal. This shows that the modulating signal increases the flexibility of the decoder, by allowing the decoder to capture information for formants and/or fricatives of the singing voice alongside the cosine functions. Nonetheless, this means that signal operations in the computed representation, like masking, will affect a greater proportion of the singing voice signal compared to typical sinusoidal functions employed by the STFT. This in turn, might not be ideal in general applications such as frequency equalization, where only the specific frequency regions have to processed in a deterministic way.

\subsection{Sinkhorn Distances Results}\label{sec:sk-distance-res}
\begin{figure}[!ht]
\centering
\begin{minipage}[!t]{0.99\columnwidth}
\includegraphics[width=0.33\columnwidth,height=4.25cm]{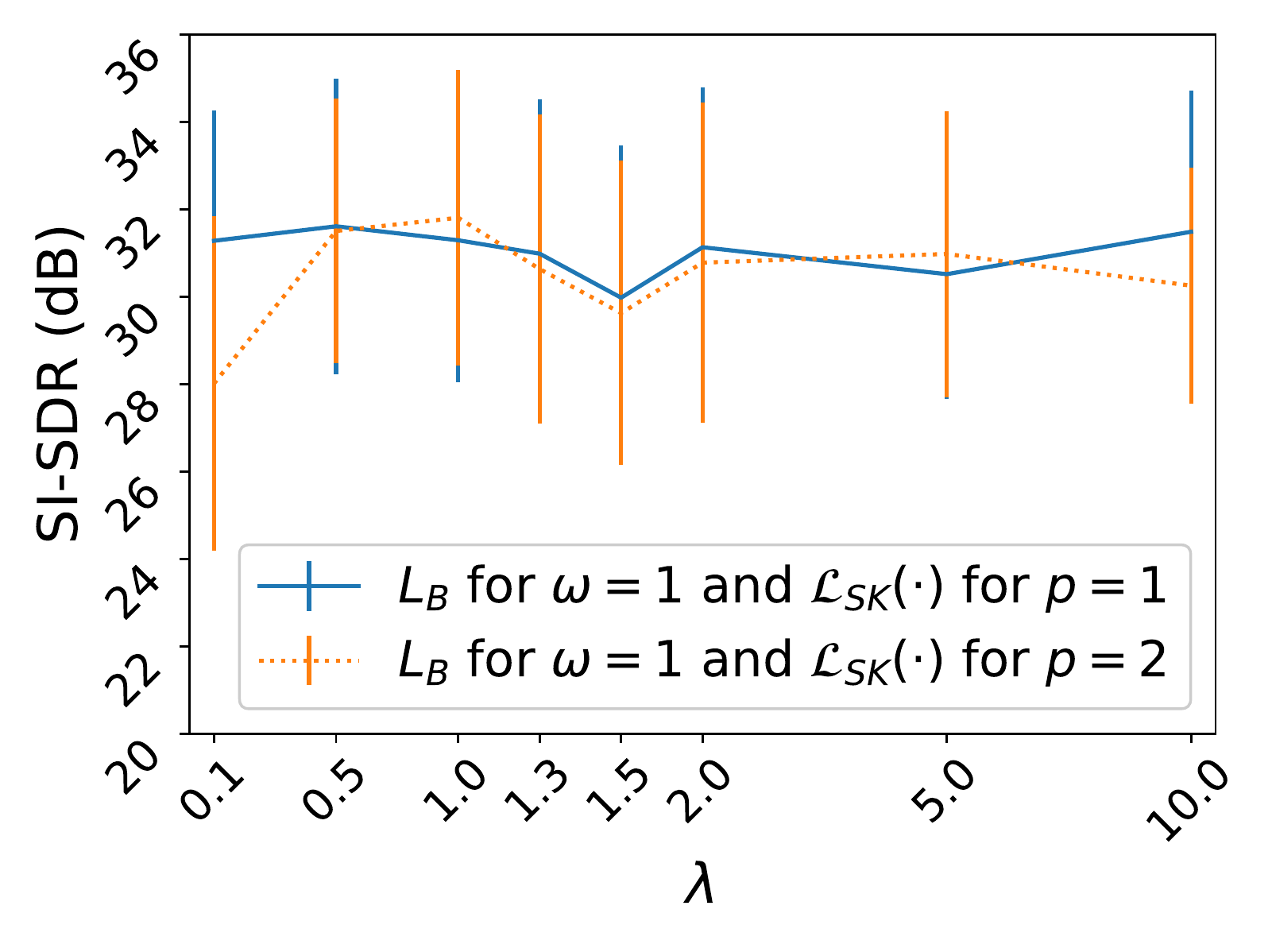}%
\includegraphics[width=0.33\columnwidth, height=4.25cm]{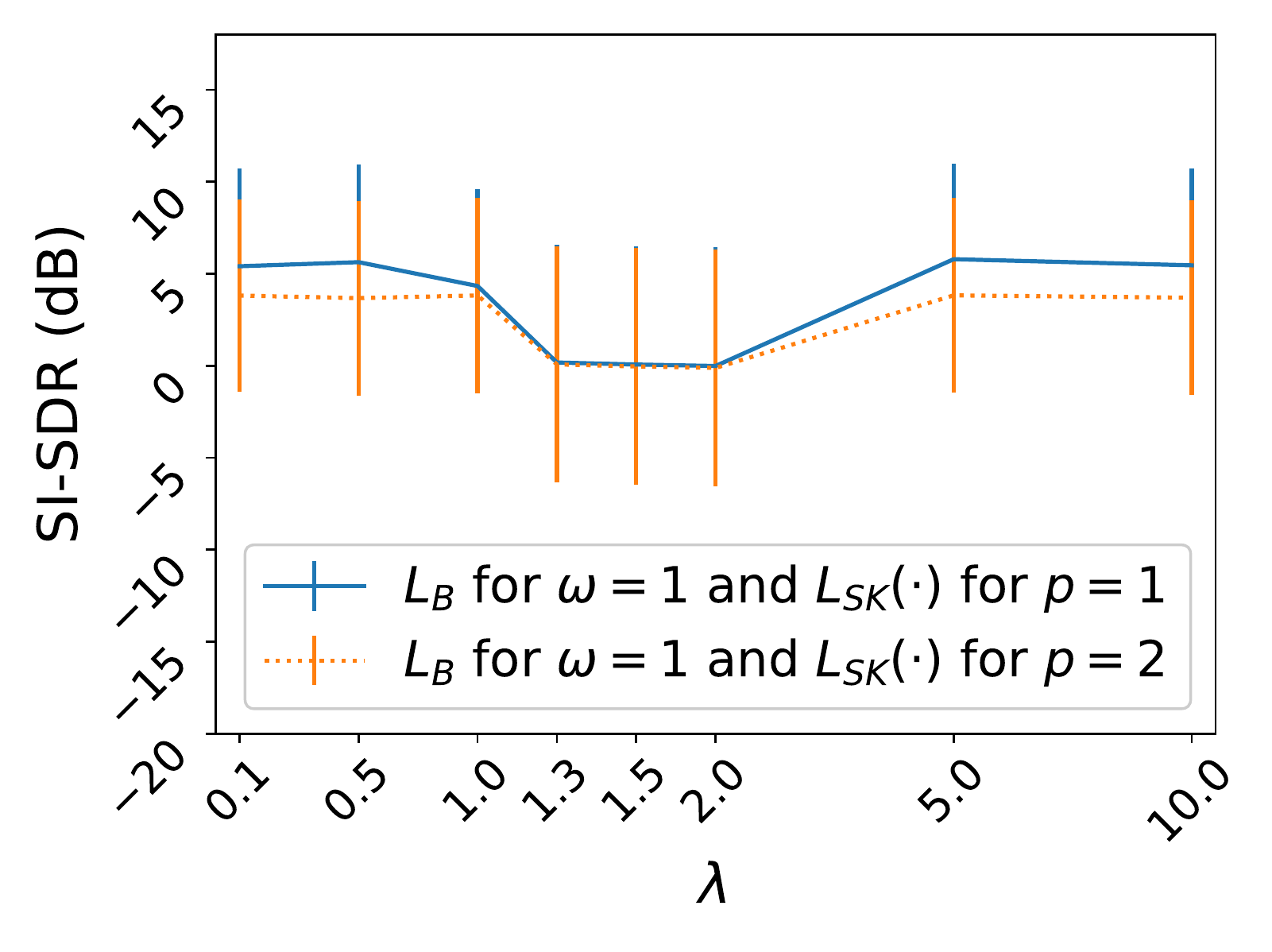}%
\centering
\includegraphics[width=0.33\columnwidth, height=4.25cm]{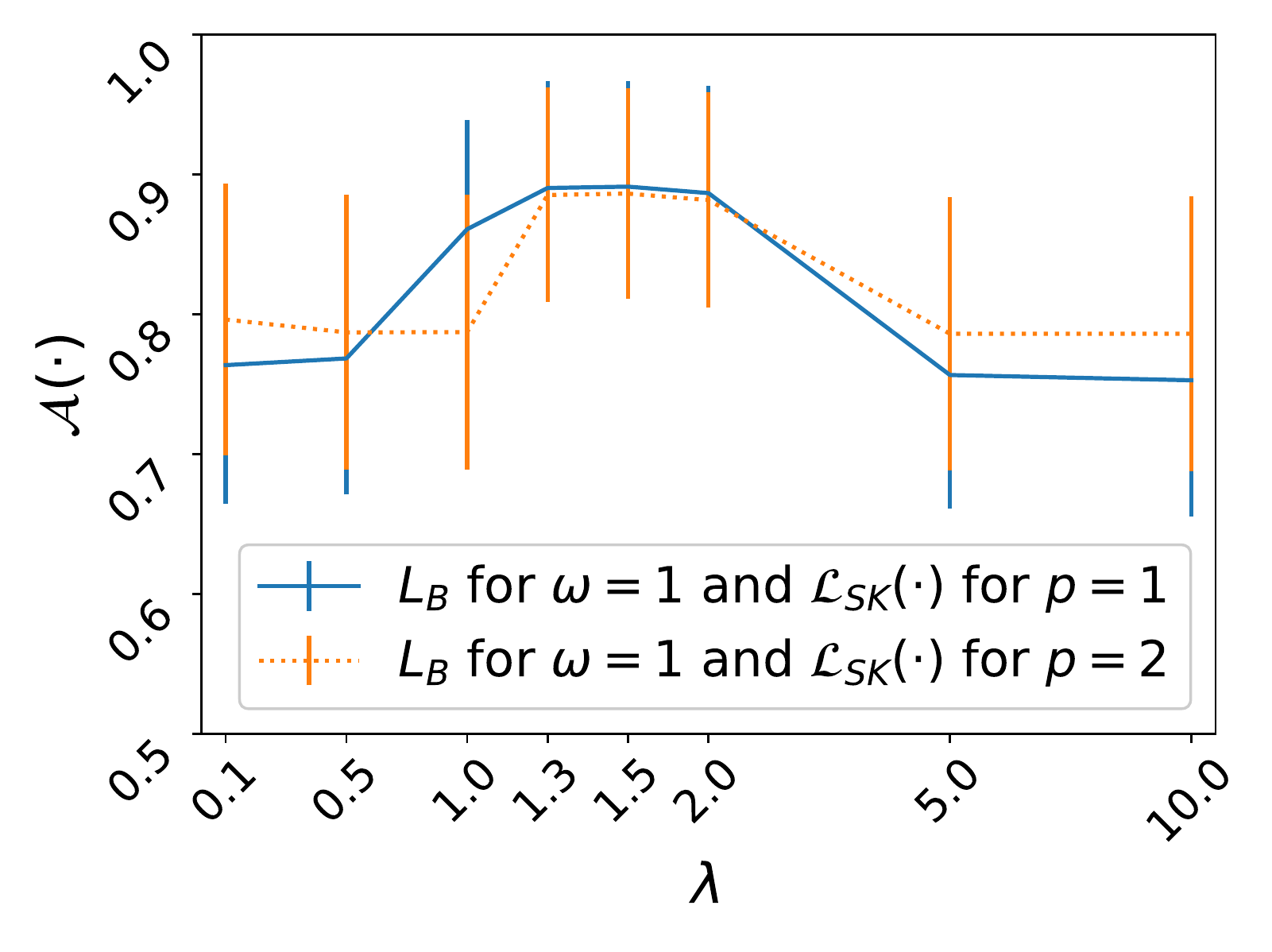}%
\end{minipage}\\
\caption{Performance evaluation of the learned representations by ${L}_{B}$ that use the Sinkhorn distances. (top-left) Reconstruction of singing voice in SI-SDR, (top-right) oracle separation performance in SI-SDR, and (bottom) additivity objective measure. Horizontal and vertical lines denote the average and the standard deviation of the performance, respectively.}
\label{fig_extra_res:sk}
\end{figure}

To complement the results from using ${L}_{B}$ that is computed using the Sinkhorn distances, Figure~\ref{fig_extra_res:sk} presents results from the objective evaluation of the learned for a greater range of hyper-parameter values. Particularly, Figure~\ref{fig_extra_res:sk} contains error plots for the following range of entropic regularization weights $\lambda \in [0.1, 0.5, 1.0, 1.3, 1.5,2.0, 5.0, 10.0]$ and for $\omega = 1.0$. To justify the choice for $p=1$ for computing the pair-wise distances matrix $\mathbf{M}$, used for computing the Sinkhorn distances, additional results  for $p=1$ and $p=2$ are illustrated in Figure~\ref{fig_extra_res:sk}. By observing Figure~\ref{fig_extra_res:sk}, two observations can be highlighted. The first observation is that the computation of the loss matrix $\mathbf{M}$ for $p=2$ leads to marginally sub-optimal results, compared to $p=1$, for nearly all $\lambda$ values and with respect to all the evaluation metrics. Specifically, the reconstruction performance for $p=1$ is better than $p=2$ by 1 dB on average across $\lambda$ values. Also, for $p=1$ an improvement of
$0.6$ dB on average with respect to the performance of separation by masking
is observed in comparison to $p=2$.
For the additivity metric, $p=2$ marginally outperforms $p=1$ for a negligible difference of $3e^{-3}$. 
These results indicate the reason why the above presented results focus on $p=1$.

Another observation from Figure~\ref{fig_extra_res:sk} is that for $\lambda > 2$ the observed separation performance dip and additivity performance peak disappear in the area of $\lambda \in [1.3, 1.5, 2.0]$. In this area the examined method performs similarly to the values of low entropy, with respect to the examined metrics. This contradicts the expectations for the effect of entropic regularization. The only explanation to this behavior is that for values $\lambda > 2$, the exponential function used in the computation of the Sinkhorn distances and is applied to $\mathbf{M}$, yields saturated values that bias the overall minimization. The unexpected effect in the minimization of the computed loss values using the Sinkhorn distances for various values of $\lambda$ is illustrated in Figure~\ref{fig_extra_res:sinkhorn} in the  Appendix.

\section{Summary}\label{ch5:sec:summary}
This manuscript presented a method for learning representations of music signals that can be particularly useful for the task of music source separation. The presented method is based on the denoising autoencoder model~\cite{DAEs:Vincent-2010} with modulated cosine functions for decoding bases, inspired by the differential digital signal processing concept~\cite{ddsp}. The benefits of the proposed method are interpretability, due to the usage of the cosine functions for decoding, non-negativity promoting energy-informative representations akin to the magnitude of the STFT, and the fact that the proposed method can be trained in an unsupervised fashion, enabling the usage of unlabeled and unpaired multi-track data.

Focusing on the important problem of singing voice separation, the proposed method was investigated for its performance in separation, additivity of the sources' representations, and the reconstruction of the singing voice signal. Furthermore, representation objective functions were examined for improving the attributes and the performance of the learned representations. Specifically, two objectives were examined, the (an-isotropic) total-variation denoising loss~\cite{tv_loss}, and the family of Sinkhorn distances with entropic regularization~\cite{sinkhorn}. The results from the experimental procedure, suggest that representations for music signals can be learned using unsupervised learning, leading representations that can be employed for the separation of singing voice by masking. In addition to this, Sinkhorn distances as an efficient computation for optimal-transportation distances, allow a flexible learning of representations in an unsupervised way, with the entropic regularization leading to sources' representations that are distinctly structured and are almost additive; attributes that are useful in music source separation. The source code is based on the Pytorch framework~\cite{NEURIPS2019_9015} and is available online\footnote{\url{https://github.com/Js-Mim/rl_singing_voice}}.

\section*{Acknowledgements}
Stylianos I. Mimilakis is supported in part by the German Research Foundation (AB 675/2-1, MU 2686/11-1). K. Drossos would like to acknowledge CSC Finland for computational resources.

\clearpage
\section*{Appendix}~\label{app} \vspace{-1cm}
\subsection*{W-Disjointness Orthogonality}
\label{app:ch-wdo}
Let $\mathbf{Y}_{j} \in \mathbb{R}^{N \times T'}$ be the representation of the source's signal $\mathbf{x}_{j} \in \mathbb{R}_{[-1,1]}^{T}$ computed using an appropriate method, where $N, T,$ and $T'$ are the number of components of frequency sub-bands of the representation, the number of time-domain samples, and the number of time-frames respectively. Furthermore, assume that the representation of the interfering source\footnote{The interfering source is the sum of all the sources in the mixture except the target source.} $\mathbf{x}_{j'} \in \mathbb{R}_{[-1,1]}^{T}$ is provided and is denoted as $\mathbf{Y}_{j'} \in \mathbb{R}^{N \times T'}$. Then, the binary mask (BM), $\mathbf{M}^{\text{BM}} \in [0,1]^{N \times T'}$ is computed using 
\[
 {\mathbf{M}_{j}}^{\text{BM}} = g \Big(|\mathbf{Y}_{j}| \oslash |\mathbf{Y}_{j'}|\Big) \text{ , and}
\]
\[g(\mathrm{y}) = 
    \begin{cases}
      1, & \text{if}\ \mathrm{y}\geq0.5 \\
      0, & \text{otherwise}
    \end{cases}
\text{ . }\]

Then the windowed disjointness orthogonality (W-DO) measure is computed as 
\begin{equation}\label{eq:w-do}
\text{W-DO} = \text{PSR} - \frac{\text{PSR}}{\text{SIR}} \text{ , }
\end{equation}
where PSR and SIR are the preserved-signal-ratio and the source-to-interference ratio computed as
\[
\text{PSR} = \frac{||\mathbf{M}_{j}^{\text{BM}} \odot |\mathbf{Y}_{j}|\,||_{1}^{2}}{||\, |\mathbf{Y}_{j}|\,||_{1}^{2}}, \, \text{SIR} = \frac{||\mathbf{M}_{j}^{\text{BM}} \odot |\mathbf{Y}_{j}|\,||_{1}^{2}}{||\mathbf{M}_{j}^{\text{BM}} \odot |\mathbf{Y}_{j'}|\,||_{1}^{2}} \text{ , }
\]
where $||\cdot||_{1}$ is the unit matrix/vector norm. From the above expressions, it can be seen that for a W-DO value of one the sources are entirely disjoint, meaning that there is not overlap between the sources in the respective representation. In contrast, a W-DO value of zero means that the sources completely overlapped and the separation of the $j$-th source by binary masking is not possible. In the latter case, the inability of separating the $j$-th source is also reflected by extremely low PSR values indicating a poor reconstruction of the source after masking.

\subsection*{Total-variation complimentary results}
\begin{figure}[!ht]
\centering
\begin{minipage}{1.\columnwidth}
\includegraphics[width=0.33\columnwidth,height=4.25cm]{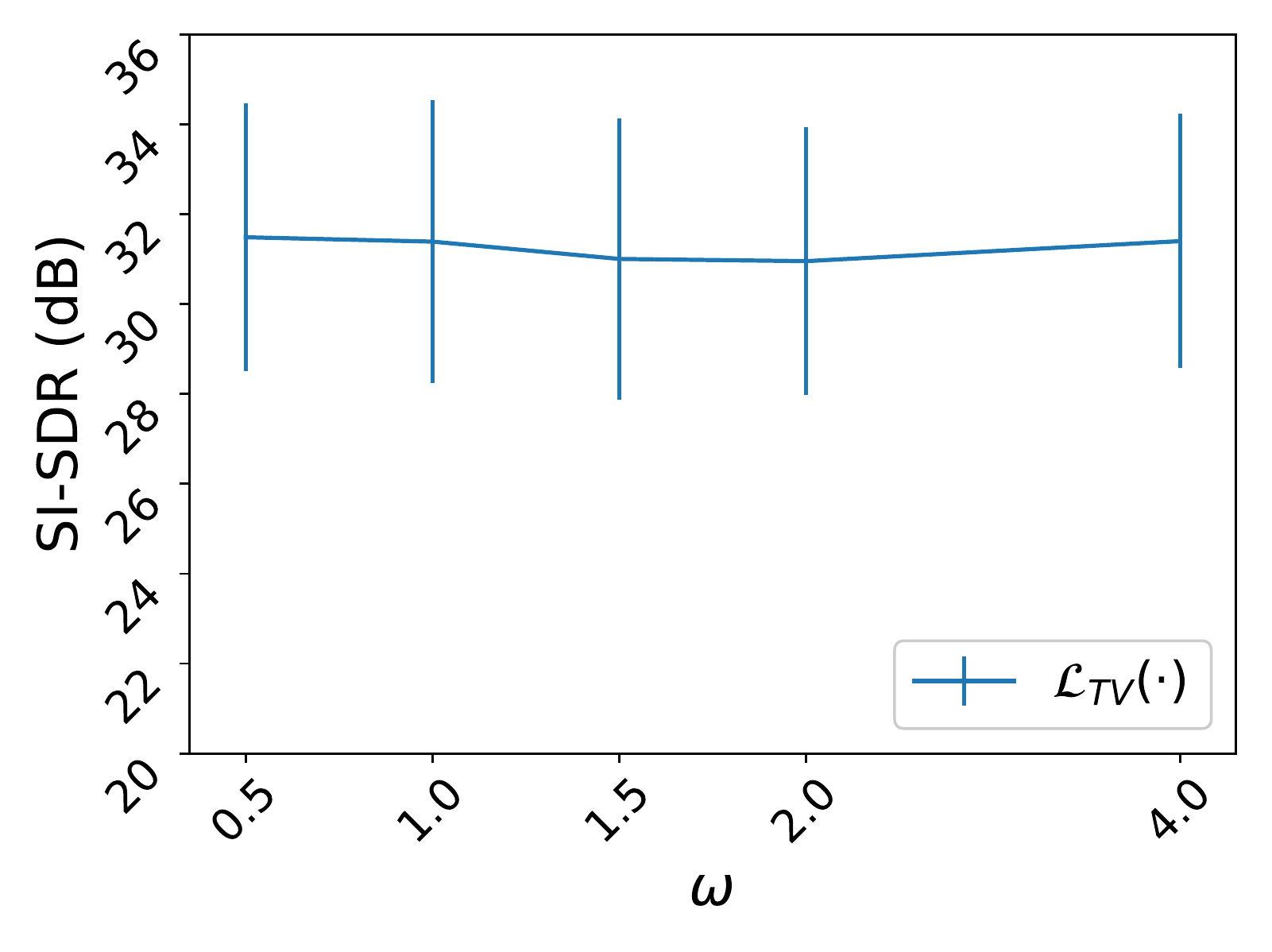}%
\includegraphics[width=0.33\columnwidth, height=4.25cm]{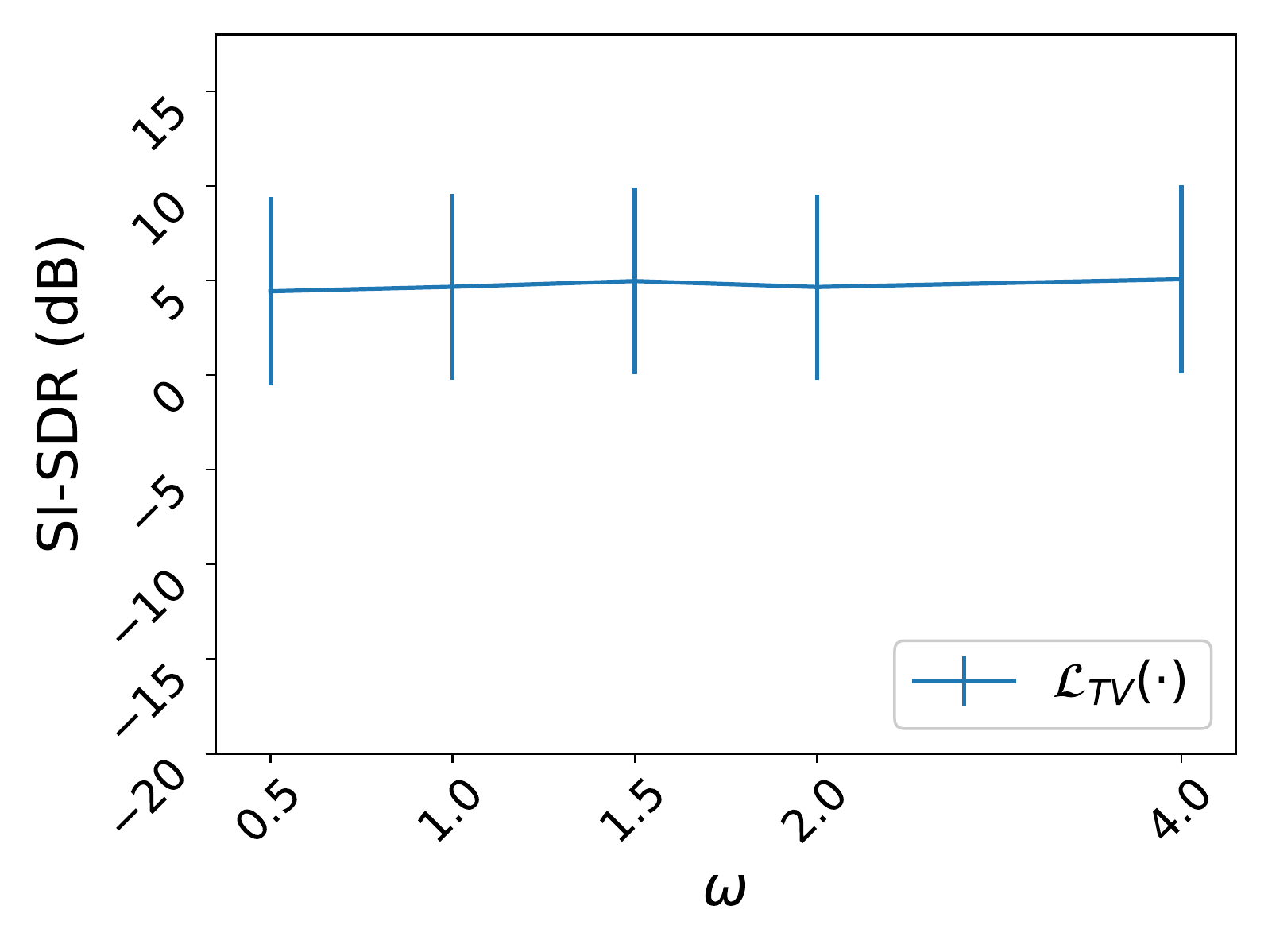}%
\includegraphics[width=0.33\columnwidth, height=4.25cm]{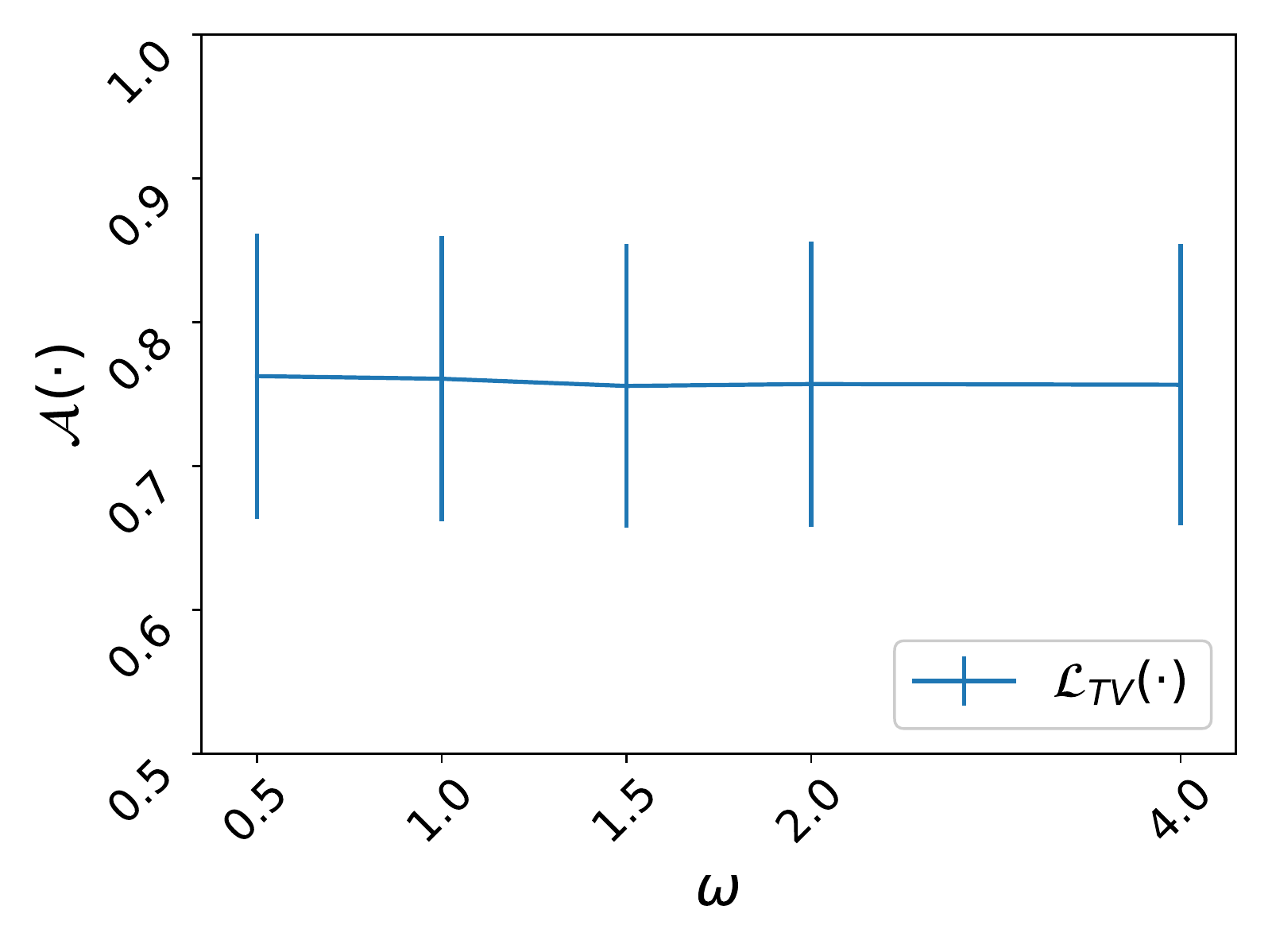}
\centering
\end{minipage}%
\caption{${L}_{A}$ using total-variation denoising ($\mathcal{L}_{\text{TV}}(\cdot)$) for various values of $\omega$:  (top-left) Reconstruction of singing voice in SI-SDR, (top-right) oracle separation performance in SI-SDR, and (bottom) additivity objective measure.}
\label{fig_extra_res:tv}
\end{figure}

In Figure~\ref{fig_extra_res:tv} complimentary results for Table~\ref{tab:res-1} are presented. Figure~\ref{fig_extra_res:tv} illustrates the obtained results using the (an-isotropic) total-variation denoising loss ($\mathcal{L}_{\text{TV}}(\cdot)$), employed in the computation of the loss termed as $L_{A}$.
\clearpage
\subsection*{Sinkhorn distances complimentary results}
\begin{figure}[!ht]
\centering
\includegraphics[width=0.55\columnwidth,keepaspectratio]{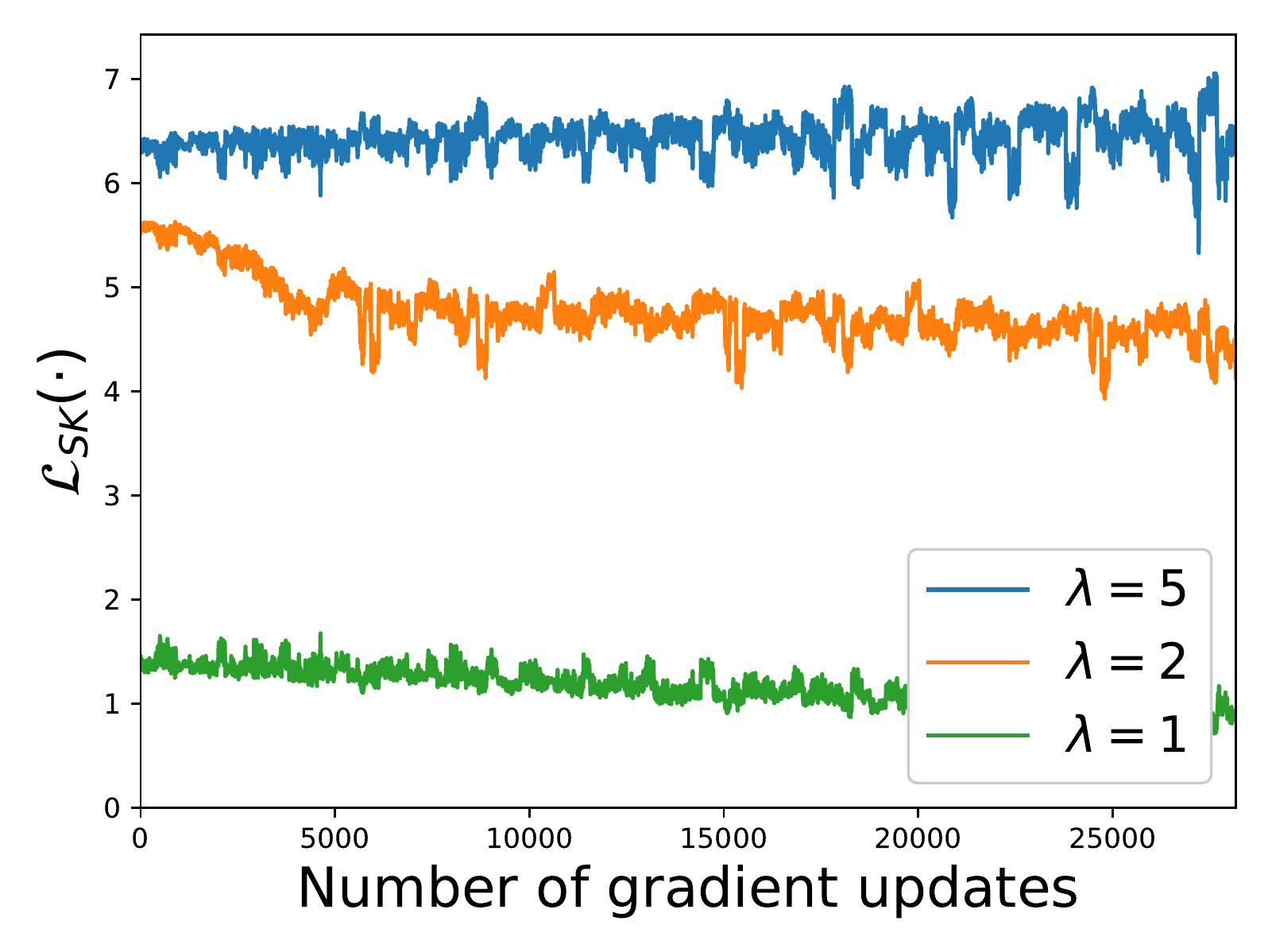}%
\caption{The output values of $\mathcal{L}_{\text{SK}}(\cdot)$ with $p=1$ and for various values for $\lambda$ that are used to compute the loss $L_{B}$. The number of gradient updates corresponds to $\sim 3$ full iterations throughout the whole training data-set. An eight-order quadratic smoothing filter with a window of $61$ gradient updates is applied to the results for clarity.}
\label{fig_extra_res:sinkhorn}
\end{figure}

Figure~\ref{fig_extra_res:sinkhorn} illustrates the output loss values of the $\mathcal{L}_{\text{SK}}(\cdot)$ for the entropic regularization values $\lambda = [1, 2, 5]$. This figure serves as complimentary experimental results that show the saturation of the computed loss values for $\lambda = [2, 5]$, aiming at explaining the unexpected behavior of entropic regularization for high $\lambda$ values discussed in Section~\ref{sec:sk-distance-res}.

\end{document}